\documentclass[
  journal=pasa,
  manuscript=research-paper, 
  year=2022,
  volume=39,
]{cup-journal}

\usepackage{microtype,siunitx,booktabs}
\usepackage{aas-macros}
\usepackage{epstopdf}
\usepackage{caption}
\usepackage{subcaption}

\usepackage{amsmath,amssymb}
\usepackage{gensymb}
\usepackage{longtable,threeparttablex}
\usepackage{physics}
\usepackage{comment}
\usepackage{hyperref} 

\usepackage{float}
\usepackage[export]{adjustbox}

\hypersetup{colorlinks,citecolor=blue,linkcolor=blue,urlcolor=blue}
\usepackage{listings}
\definecolor{codegreen}{rgb}{0,0.6,0}
\definecolor{codered}{rgb}{0.6,0,0}
\definecolor{codeblue}{rgb}{0,0,0.6}
\definecolor{codegray}{rgb}{0.5,0.5,0.5}
\definecolor{codepurple}{rgb}{0.58,0,0.82}
\definecolor{codeorange}{RGB}{200, 112, 60}
\definecolor{backcolour}{rgb}{0.95,0.95,0.95}

\lstdefinestyle{mystyle}{
    backgroundcolor=\color{backcolour},   
    commentstyle=\color{codegreen},
    keywordstyle=\color{codepurple},
    numberstyle=\tiny\color{codegray},
    stringstyle=\color{codeorange},
    basicstyle=\ttfamily\footnotesize,
    breakatwhitespace=false,         
    breaklines=true,                 
    captionpos=b,                    
    keepspaces=true,                 
    numbers=left,                    
    numbersep=5pt,                  
    showspaces=false,                
    showstringspaces=false,
    showtabs=false,                  
    tabsize=4
}

\def\dmu{\ensuremath{\rm pc\,cm^{-3}}}
\def\lmu{\ensuremath{\rm mJy\,kpc^{2}}}
\def\rmu{\ensuremath{\rm rad\,m^{-2}}}
\def\TBhr{\ensuremath{\rm TB\,hr^{-1}}}
\def\accu{\ensuremath{\rm m\,s^{-2}}}
\def\us{\ensuremath{\mu{\rm s}}}  
\def\arcmin{\ensuremath{\prime}}
\def\arcsec{\ensuremath{{\prime\prime}}}
\def\mJy{\ensuremath{\rm mJy}}

\def\KperJy{${\rm K\,Jy^{-1}} $\,}
\def\deg{\ensuremath{{^{\circ}}}}
\def\sqdeg{${\rm deg^2}$\,}
\def\sqm{$ {\rm m^2} $\,}
\def\sqdeghr{${\rm deg ^ {2} \, hr}$\,}
\def\sqdegphr{${\rm deg ^ {2} \, hr^{-1}}$\,}
\def\Tsky{\ensuremath{T_{\rm sky}}}
\def\Tsys{\ensuremath{T_{\rm sys}}}
\def\Trecv{\ensuremath{T_{\rm recv}}}
\def\Smin{\ensuremath{S _{\rm min}}}

\def\Npol{\ensuremath{n _{\rm pol}}}
\def\Tobs{\ensuremath{t _{\rm obs}}}
\def\Bobs{\ensuremath{B _{\rm obs}}}

\def\Aeff{\ensuremath{A _{\rm eff}}}
\def\Weff{\ensuremath{W _{\rm eff}}}  
\def\tautot{\ensuremath{\tau _{\rm tot}}}  
\def\tausamp{\ensuremath{\tau _{\rm samp}}}  
\def\tauchan{\ensuremath{\tau _{\rm chan}}}  
\def\tausub{\ensuremath{\tau _{\rm sub}}}  
\def\taudm{\ensuremath{\tau _{\rm dm}}}  

\newcommand{\psrone}{J0036$-$1033}
\newcommand{\psrtwo}{J0026$-$1955}
\newcommand{\psrthree}{J1002$-$2036}

\newcommand{\cdmtpsr}{J0952$-$0607}
\newcommand{\hitrun}{J1757$-$1854}

\title[The SMART pulsar survey]{The Southern-sky MWA Rapid Two-metre (SMART) pulsar survey - I. 
Survey design and processing pipeline}

\author{N. D. R. Bhat}
\affiliation{International Centre for Radio Astronomy Research, Curtin University, Bentley, WA 6102, Australia}
\author{N. A. Swainston}
\affiliation{International Centre for Radio Astronomy Research, Curtin University, Bentley, WA 6102, Australia}
\author{S. J. McSweeney}
\affiliation{International Centre for Radio Astronomy Research, Curtin University, Bentley, WA 6102, Australia}
\author{M. Xue}
\affiliation{National Astronomical Observatories, Chinese Academy of Sciences, Datun Road, Chaoyang District, Beijing 100101, China}
\author{B. W. Meyers}
\affiliation{International Centre for Radio Astronomy Research, Curtin University, Bentley, WA 6102, Australia}
\alsoaffiliation{Department of Physics \& Astronomy, University of British Columbia, 6224 Agricultural Road, Vancouver, BC V6T 1Z1, Canada}
\author{S. Kudale}
\affiliation{National Centre for Radio Astrophysics, Tata Institute of Fundamental Research, Pune 411 007, India}
\author{S. Dai}
\affiliation{Western Sydney University, Locked Bag 2751, Penrith South DC, NSW 1797, Australia}
\author{S. E. Tremblay}
\affiliation{National Radio Astronomy Observatory, 1003 Lopez Road, Socorro NM, 87801, USA}
\author{W. van Straten}
\affiliation{Institute for Radio Astronomy \& Space Research, Auckland University of Technology, Private Bag 92006, Auckland 1142, New Zealand}
\author{R. M. Shannon}
\affiliation{Centre for Astrophysics and Supercomputing, Swinburne University of Technology, P.O. Box 218, Hawthorn, VIC 3122, Australia}
\author{K. R. Smith}
\affiliation{International Centre for Radio Astronomy Research, Curtin University, Bentley, WA 6102, Australia}
\author{M. Sokolowski}
\affiliation{International Centre for Radio Astronomy Research, Curtin University, Bentley, WA 6102, Australia}
\author{S. M. Ord}
\affiliation{CSIRO Astronomy and Space Science, PO Box 76, Epping, NSW 1710, Australia}
\author{G. Sleap}
\affiliation{International Centre for Radio Astronomy Research, Curtin University, Bentley, WA 6102, Australia}
\author{A. Williams}
\affiliation{International Centre for Radio Astronomy Research, Curtin University, Bentley, WA 6102, Australia}
\author{P.~J.~Hancock}
\affiliation{Curtin Institute for Computation, Curtin University, GPO Box U1987, Perth, 6845, WA, Australia}
\author{R.~Lange}
\affiliation{Curtin Institute for Computation, Curtin University, GPO Box U1987, Perth, 6845, WA, Australia}
\author{J.~Tocknell}
\affiliation{AAO Macquarie, Macquarie University, NSW, Australia}
\author{M.~Johnston-Hollitt}
\affiliation{Curtin Institute for Computation, Curtin University, GPO Box U1987, Perth, 6845, WA, Australia}
\author{D.~L.~Kaplan} 
\affiliation{Department of Physics, University of Wisconsin--Milwaukee, WI 53201, USA}
\author{S. J. Tingay}
\affiliation{International Centre for Radio Astronomy Research, Curtin University, Bentley, WA 6102, Australia}
\author{M.~Walker}
\affiliation{International Centre for Radio Astronomy Research, Curtin University, Bentley, WA 6102, Australia}

\doi{XX.XXXX/pasa.XXXX.XX}

\received {dd Mmm YYYY}
\revised  {dd Mmm YYYY}
\accepted {dd Mmm YYYY}
\published{dd Mmm YYYY}


\begin{document}


\begin{abstract}
We present an overview of the Southern-sky MWA Rapid Two-metre (SMART) pulsar survey that exploits the Murchison Widefield Array's large field of view and voltage capture system to survey the sky south of 30\deg{} in declination for pulsars and fast transients in the 140-170\,MHz band. The survey is enabled by the advent of the Phase II MWA's compact configuration, which offers an enormous efficiency in beam-forming and processing costs, thereby making an all-sky survey of this magnitude tractable with the MWA. Even with the long dwell times employed for the survey (4800\,s), data collection can be completed in $<$100 hours of telescope time, while still retaining the ability to reach a limiting sensitivity of $\sim$2-3\,mJy (at 150\,MHz, near zenith), which is effectively 3-5 times deeper than the previous-generation low-frequency southern-sky pulsar survey, completed in the 1990s. 
Each observation is processed to generate $\sim$5000-8000 tied-array beams that tessellate the full $\sim$610\,\sqdeg field of view (at 155\,MHz), which are then processed to search for pulsars. The voltage-capture recording of the survey also allows a multitude of {\it post hoc} processing options including the reprocessing of data for higher time resolution and even exploring image-based techniques for pulsar candidate identification. Due to the substantial computational cost in pulsar searches at low frequencies, the survey data processing is undertaken in multiple passes: in the first pass, a shallow survey is performed, where 10 minutes of each observation is processed, reaching about one-third of the full search sensitivity. 
Here we present the system overview including details of ongoing processing and initial results. Further details including first pulsar discoveries and a census of low-frequency detections are presented in a companion paper. 
Future plans include deeper searches to reach the full sensitivity and acceleration searches to target binary and millisecond pulsars. 
Our simulation analysis forecasts $\sim$300 new pulsars upon the completion of full processing.
The SMART survey will also generate a complete digital record of the low-frequency sky, which will serve as a valuable reference for future pulsar searches planned with the low-frequency Square Kilometre Array. 
\end{abstract}


\section{INTRODUCTION }
\label{sec:intro}

Even after five decades of productive research, pulsars continue to enable us to push the frontiers of  physics and astrophysics. These compact dense stars harbour physical conditions that are non-existent elsewhere in the universe (e.g., ultra-strong gravitational and magnetic fields and supra-nuclear matter densities), which make them invaluable tools for studying extreme physics.  They are arguably amongst the most widely-exploited astrophysical objects, with applications ranging from probing the state of ultra-dense matter to testing strong-field gravity \citep[e.g.,][]{demorest2010,kramer2006,vanstraten2001}, and from probing micro-arcsecond structure and turbulence in the interstellar medium (ISM) to complex stellar evolutionary scenarios  \citep[e.g.,][]{bhat2004,archibald2009,bailes2011}. The phenomenal impact and high-profile scientific applications (e.g., pulsar timing arrays for the detection of nanohertz-frequency gravitational waves) has elevated pulsar science to the ranks of a key science for the Square Kilometre Array \citep[SKA; e.g.,][]{keane2015,janssen2015,shao2015}.

The backbone that enables this is the net result of a series of large pulsar surveys conducted over the past five decades \citep[e.g.,][]{manchester2001,cordes2006,keith2010,stovall2014}.
Invariably, most of them involved tessellating large parts of the sky of the instrument and recording data at high time and frequency resolutions (i.e., large data rates) and performing sensitive searches over the vast parameter space that is practically feasible. Many of them were prompted by the advent of new instrumentation or technology, and often exploited the computing affordable at the time. They have also proven invariably rewarding in the longer term, and often yielded a substantial increase in the pulsar population. For instance, the Molonglo pulsar survey in the 1970s found 150 pulsars, practically doubling the known pulsar population at the time \citep{manchester1978}, while the Parkes multibeam survey from the 1990s \citep{manchester2001} found 742 pulsars, and discovered exotica such as the double pulsar system J0737$-$3039A/B and the eccentric neutron star-white dwarf binary J1141$-$6545, both of which have proven to be unique laboratories for testing general relativity and alternate theories of gravity
\citep{kramer2006,bhat2008,vivek2020,kramer2021}. This success led to next-generation multibeam surveys at Parkes and Arecibo, and more recently with the Five-hundred-meter Aperture Spherical radio Telescope (FAST).
Already these have collectively discovered 600 pulsars to date. The landmark discovery of fast radio bursts (FRBs) in the Parkes high-time resolution radio universe (HTRU) survey \citep{thornton2013} even opened up an entirely new field of research. Large pulsar surveys have a proven track record of their ability to return significant scientific dividends in the long run, with the majority of the discoveries and spin-off science emerging from follow-up processing over the years. 

These multi-beam surveys have largely been at frequencies $\gtrsim 1$\,GHz. The past decade also witnessed a number of successful low-frequency pulsar surveys, most of which were prompted by the advent of  new-generation low-frequency facilities (e.g., Low Frequency Array; LOFAR), or  new receivers or pulsar instrumentation at the more traditional facilities such as the Green Bank Telescope (GBT) and the Giant Metre-wave Radio Telescope (GMRT). Notable among these are the drift-scan surveys with the Arecibo Telescope and GBT, and the ongoing surveys at the GMRT and GBT. The drift-scan surveys, in the 300-350\,MHz range, despite their non-traditional nature, have led to $>100$ pulsar discoveries, while the highly successful Green Bank Northern Celestial Cap (GBNCC) survey has, to date, found 160 pulsars. The net tally from the low-frequency surveys of the past decade alone is $>$400 pulsars, including 73 pulsars by the LOFAR Tied-Array All-Sky (LOTAAS) survey \citep{sanidas2019}. Additionally, targeted searches have been undertaken toward unidentified Fermi gamma-ray sources (mostly at low frequencies), leading to $>$80 pulsars \citep[and references therein]{deneva2021}. The LOTAAS survey, the processing of which is still ongoing, also discovered the longest-period (23.5\,s) pulsar known until recently \citep{tan2018b}, when a 76-s pulsar was discovered with MeerKAT \citep{caleb2022}. In essence, surveys at low frequencies have proven to be highly effective, particularly in uncovering the local population of pulsars, and mapping out the high-Galactic latitude ($b$) parts of the sky. 

Surveys at low frequencies offer several benefits but they also have their limitations. An appealing factor is the generally steep spectral nature of most radio pulsars, where the flux density at frequency $\nu$ is $ S_\nu \propto \nu^{\alpha}$, where $\alpha$ is the spectral index. The spectral index is known to vary over a wide range for pulsars, $-4 \lesssim \alpha \lesssim 0$, but the average spectral index $\langle \alpha \rangle = -1.6 \pm 0.03 $ for long-period pulsars \citep{jankowski2018}, and is somewhat steeper ($\langle \alpha \rangle = -1.9 \pm 0.1 $) for MSPs \citep{toscano1998,dai2015}, with a 1-$\sigma$ dispersion of $\sim$1. While this suggests most pulsars are significantly brighter at low frequencies, this is more than offset by the even steeper dependence of the sky background noise ($ \Tsky \propto \nu^{-2.55}$). The sky background is also highly direction-dependent and is typically significantly reduced toward higher Galactic latitudes.  The main benefit is the inherently larger fields-of-view of the low-frequency telescopes, which  substantially increase the efficiency in telescope time required and hence the time for completion of large surveys. 

Amongst the multitude of other considerations are interstellar medium (ISM)  propagation effects, which tend to majorly influence low-frequency pulsar searches; the most familiar (and significant) one  is the dispersion that manifests as frequency-dependent time delays in arrival times $\Delta t \propto {\rm DM} \, \nu^{-2}$, where the dispersion measure (${\rm DM}$) is
the line-of-sight integral of the electron density $n_e$. This non-linear, inverse dependence in frequency implies very large delays at low frequencies ($\lesssim$200\,MHz); e.g., a pulsar with a DM = 100\,\dmu{} will have its signal spread over ${\sim}7.5$\,s in observations made over a 30 MHz band centred at 150\,MHz, as opposed to $\lesssim 0.1\,$s across a similar (i.e., 20\%) fractional bandwidth around 1.4\,GHz. Circumventing this necessitates much finer frequency resolution ($\Delta \nu$) so the residual dispersive smearing across the finite channel width can be minimised, and consequently requires many more channels across the recording bandwidth, and hence a much larger data rate and substantial processing needs.

The other significant effect is pulse broadening resulting from multipath propagation as a consequence of scattering in the ISM,
the characteristic time for which is a nonlinear function of both DM and frequency, i.e., $\tau _d \propto {\rm DM}^{-2.2}\, \nu^{-4.4}$,  under the assumption of a pure Kolmogorov form of electron density spectrum \citep{cwb85}. This poses a significant limitation in low-frequency pulsar searches, especially when the pulse broadening time exceeds the pulsar's spin period, i.e., $\tau _d \gtrsim P$, as it results in a significant degradation or even a loss of sensitivity to periodic emission. As with the sky background, scatter broadening is also highly line-of-sight dependent; it is much larger in the plane, or toward the Galactic Centre, compared to  high-$|b|$ sight lines. Empirical relations exist to guide expected broadening times as a function of DM and frequency \citep[e.g.,][]{bhat2004,geyer2017}, and can be used to guide the observing/search strategies, e.g., $\tau _d \gtrsim 100$\,ms at DM $\gtrsim$300\,\dmu, for a line of sight as far off as $|b|\sim 5\deg$ and $\sim 30\deg$ away from the Galactic Centre (GC) in longitude. This implies, at low frequencies,  the search volume is largely limited to a few kpc in the plane.  However, this is not a serious limitation at higher Galactic latitudes, 
where the DM tends to saturate at $\sim$20-50\,\dmu{} for $|b| > 15\deg$. In other words, the higher survey speeds of low-frequency surveys can be optimally exploited for covering high-$|b|$ parts of the sky, without compromising detection sensitivity. 

Yet another relevant ISM effect, especially at low frequencies, is the modulation of apparent pulsar intensities due to scintillation effects. As with the pulse broadening, the observable effects strongly depend on frequency and the line of sight, as it is essentially another manifestation of multipath propagation. For relatively nearby pulsars (DM $\lesssim$50\,\dmu) this often manifests as rapid (and very large) modulations in both time and frequency with characteristic scales in the range $\sim$0.1-5\,MHz and $\sim$1-100\,min at $\sim$150\,MHz; this  is diffractive scintillation \citep[e.g.,][]{barney1990}. Refractive scintillation also leads to intensity modulations, but on much longer timescales of days to weeks (at low frequencies), and the observed variations in mean flux densities can be as much as by a factor $\sim$5-6 for low to moderate DM pulsars \citep[e.g.,][]{bell2016,bhat2018}. From the 
perspective of candidate detection in low-frequency searches, this  sometimes  results in fortuitous brightening (or inauspicious dimming) of pulsars, which 
provides the opportunity to detect pulsars that were missed earlier (e.g., owing to scintillation dimming), or to detect a pulsar that 
might be below the sensitivity limit of a survey. 
This further strengthens the case for low-frequency surveys.

Despite these challenges, pulsars were originally discovered at low frequencies (at 81.5\,MHz; \citealt{hewish1968}) and much of the early years of pulsar astronomy were focused at low frequencies. The eventual quest to find them in large numbers and timing them at high precision pushed much of pulsar astronomy (searches and timing in particular) to frequencies $\gtrsim$1\,GHz. However, the advent of several low-frequency telescopes over the past decade and advances in affordable high-performance computing are effectively leading to a resurgence of low frequency astronomy including large sky surveys, many of which are conducted at frequencies $\lesssim$500\,MHz.

The success of these northern surveys strongly motivates an all-sky pulsar survey with the Murchison Widefield Array (MWA) that operates in the 80-300\,MHz range  in the Southern Hemisphere. The MWA, which was originally built as an array of 128 tiles (where each tile is a 4$\times$4 dipole array) with a maximum baseline of 3\,km, is also Australia's precursor for the low-frequency SKA (i.e., SKA-Low; \citealt{tingay2013}). Even though the MWA was not initially designed for pulsar science, the eventual addition of a voltage capture system (VCS; \citealt{tremblay2015}) and the development of software-defined instrumentation (for offline processing) equipped it as a pulsar-capable facility. Notwithstanding the limitations of large data rates (28\,\TBhr) and the associated data management/processing challenges, the VCS has been exploited for wide-ranging science from studies of millisecond pulsars to sporadic emission from pulsars \citep[e.g.][]{bhat2016,meyers2018,kaur2019}, and from investigating the pulsar emission physics to studying propagation effects caused by the interstellar medium \citep[e.g.][]{mcsweeney2017,kaur2022}. The progress in this area, along with the array's upgrade to Phase II \citep{wayth2018}, whereby a compact configuration of 128 tiles within 300\,m was possible on a semi-regular basis, has made all-sky pulsar searches tractable with this telescope. 

The SMART survey described in this paper has two main objectives: (1) performing sensitive searches for pulsars and fast transients in the sky south of $+30\deg$ in declination at 140-170\,MHz; and (2) mapping the sky for low-frequency detection of already known pulsars in the southern sky. The main novelty of the survey is the use of a voltage-capture mode for data recording (as opposed to the filterbank data format that has been adopted for all past and ongoing surveys), and hence an astonishingly high survey speed for data collection, i.e., $\sim$500\,\sqdegphr in 100-$\mu$s/10-kHz resolutions). However, the computational cost of processing  (i.e., beamforming and searching) are substantial at low frequencies, and thus drive the feasible strategies for data processing, especially at early stages. 

With the large survey speed substantially reducing the demand for telescope time for survey completion, longer dwell times become affordable, which also increases the sensitivity to the detection of sporadic or intermittent class of objects such as rotating radio transients (RRATs; \citealt{rrats2006}), intermittent or state-switching pulsars, extreme nullers etc. \citep[e.g.,][]{kerr2014} among the classes of radio-emitting neutron stars, and even the enigmatic fast radio bursts (FRBs; e.g., \citealt{thornton2013}). The detectability all of these transient class of objects is dictated by the ``on-sky'' time metric $\Sigma = \Omega T$ where $\Omega$ is the instantaneous FoV and $T$ is the time spent on sky (dwell time in the case of an all-sky survey). Following the discussion in \citet{sanidas2019} in the context of LOTAAS, $\Sigma _{\rm SMART}$ = 52,735\,\sqdeghr, which is a factor of two more than that of LOTAAS for which $\Sigma _{\rm LOTAAS}$ = 23,400\,\sqdeghr (at 135\,MHz), and indeed much larger than $\Sigma _{\rm GBNCC}$ = 1430\,\sqdeghr, $\Sigma _{\rm GHRSS}$ = 835\,\sqdeghr and $\Sigma _{\rm AO327}$ = 132\,\sqdeghr (all at 300-350\,MHz). 

Here we present an overview of the Southern-sky MWA Rapid Two-metre (SMART) pulsar survey. In \S~\ref{sec:surveydescription} we outline the main science goals, and describe the observing strategy adopted for sky tessellation. 
Procedures for data processing and analysis are described in \S~\ref{sec:dataprocessingandanalysis}, and the strategies for confirmation and initial
follow-up in \S~\ref{sec:confirmationandfollowup}. 
In \S~\ref{sec:simulations} we describe the survey simulations and the expected yield. 
Future processing plans are outlined in \S~\ref{sec:discussionandfutureplans}, followed by a summary in \S~\ref{sec:summary}. 


\section{Survey Description }
\label{sec:surveydescription}

\subsection{Science goals and Motivation }
\label{sec:sciencegoals}

The broader goals of the SMART survey are similar to most other large sky surveys, i.e., exploring the new parameter space that is opened by a leap in instrumentation, technology, or sensitivity and to uncover a large population of previously undetected pulsars.
The fact that the currently known pulsar population ($\sim$3300, cf. the ATNF pulsar catalogue\footnote{\url{https://www.atnf.csiro.au/research/pulsar/psrcat/}} v1.67; \citealt{manchester2005}) represents only a small fraction ($\lesssim$ 10\%) of the total expected (i.e. beamed in our direction)
Galactic population \citep[e.g.,][and references therein]{keane2015} strongly motivates such large sky surveys. Indeed, conducting a full Galactic census of pulsars is a high-priority science objective for the SKA. Further, given the number of broader questions surrounding the neutron-star population (e.g., birth rates, and comparison with rates of supernovae), the detectable pulsar population is largely guided by the known population of pulsars at any given time. It is therefore imperative to explore every possible avenue and steadily refine our knowledge of pulsar population. Furthermore, the detection prospects of pulsars in a given frequency band strongly depends on the emission and propagation properties at those frequencies; however, the current forecast of a detectable population in the SKA-Low band is largely guided by the pulsar population uncovered by high-frequency surveys. 

Obtaining a large body of measurements such as DM, scattering and Faraday rotation, by using pulsars as probes of the ISM, will also enable mapping out the distribution of magneto-ionic (and turbulent) plasma in the Galaxy, which is steadily refined with a larger sample of measurements \citep[e.g.,][]{ne2001,bhat2004,deller2016,ymw16}. 

Finally, an underlying goal of any large-sky pulsar survey is to discover exotic objects; while it is hard to design any particular survey specifically for this, historical examples are abundant, e.g., the discovery of the double pulsar in the Parkes multibeam (PMB) survey \citep{lyne2004}, the 23.5-second period pulsar in LOTAAS \citep{tan2018b}, and the transitional millisecond pulsar (MSP) in the Arecibo drift-scan survey \citep{archibald2009}. All such broader and high-profile objectives are certainly applicable for the SMART survey.

\begin{figure*}[p]
\centering
\includegraphics[max size={\textwidth}{\textheight}]{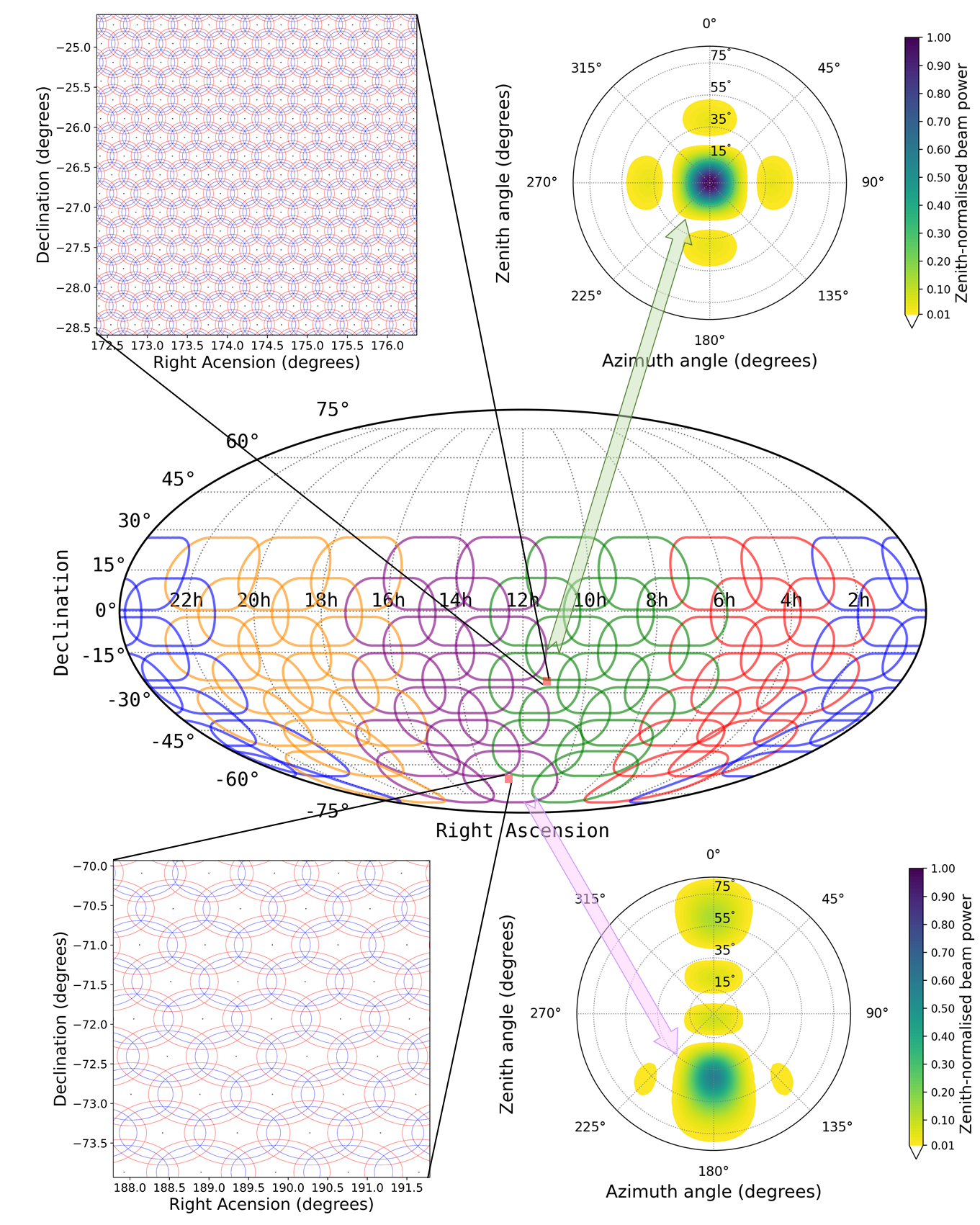}
\caption{
Sky tessellation of the SMART survey. The left panels show beam tiling patterns for two select pointings: top one a near-zenith pointing ($\delta=-28\deg$), the bottom one a far southern pointing ($\delta = -70\deg$). The number of tied-array beams vary from $\sim$6000 to $\sim$8000 from near-zenith to far-zenith pointings, and the beam shape becomes elliptical at large offsets from the zenith. The size of the circle/ellipse indicates half power tied-array beam size; the red and blue circles correspond to the low and high ends of the SMART band (140-170\,MHz). The right panels show the primary beam response for the same declination pointings, at the central frequency of 155\,MHz.
}
\label{fig:skytessellation}
\end{figure*}

\begin{table*}[!t]
\caption{Parameters of large pulsar surveys over the past decade} \label{tab:surveys} \begin{tabular}{llccccccl}
\hline
  \multicolumn{1}{c}{Survey} &
  \multicolumn{1}{c}{Telescope} &
  \multicolumn{1}{c}{Frequency Band} &
  \multicolumn{1}{c}{Sky coverage} &
  \multicolumn{1}{c}{Time resolution} &
  \multicolumn{1}{c}{Frequency resolution} &
  \multicolumn{1}{c}{ Dwell time } & 
    \multicolumn{1}{c}{ $ S_{\rm min} $ $^{\dagger} $ } &
  \multicolumn{1}{c}{ Reference } 
\\
  \multicolumn{1}{c}{ } &
   \multicolumn{1}{c}{ } &
  \multicolumn{1}{c}{(MHz)} &
  \multicolumn{1}{c}{ } &
  \multicolumn{1}{c}{ ($\mu$s)} &
  \multicolumn{1}{c}{(kHz)} &
  \multicolumn{1}{c}{(s)} &
   \multicolumn{1}{c}{(mJy)} &
   \multicolumn{1}{c}{ }
\\
\hline  
LOTAAS  &  LOFAR &  119-151     & $\delta > 0\deg $ & 491.52 & 12.21 & 3600 & 1-2 & SCB+19 \\
SMART   &   MWA & 140-170     & $\delta < +30\deg $ & 100 & 10 & 4800 & 2-3 & This work \\
GBNCC   &  GBT &  300-400     & $\delta > -40\deg $  &  81.92 & 24 & 120 & 1.1 & SLR+14 \\
GHRSS   &  GMRT &  306-338     & $ -40\deg > \delta > -54\deg $ & 30.72-61.44 & 15.625-31.25 & 900, 1200 & 1.0 & BCM+16 \\ 
HTRU    &  Parkes &  1182-1522   & $ \delta < +30\deg $ & 64 & 390 & 240, 540, 4200 & 0.2-0.6 & KJvS+10 \\ 
GPPS    &  FAST &  1100-1500   & $ -10\deg < b < +10\deg$ & 49.152 & 244.14 & 300 & 0.005 & HWW+21 \\ 

\hline
\end{tabular}
\begin{tablenotes}
\item[$^{\dagger}$] Minimum detectable flux density for a 10-$\sigma$ detection, for long-period pulsars ($P \gtrsim 0.1$\,s), with small duty cycle ($W/P\sim0.05$), and at DMs $\lesssim$50\,\dmu. 
\item[Notes:] Survey description reference -- SCB+19: \citet{sanidas2019} for LOTAAS; SLR+14: \citet{stovall2014} for GBNCC; BCM+16: \citet{bhaswati2016} for GHRSS; KJvS+10: \citet{keith2010} for HTRU; HWW+21: \citet{han2021} for GPPS
\end{tablenotes}
\end{table*}

The SMART pulsar survey also perfectly complements ongoing northern-sky surveys in sky and frequency coverage (Table~\ref{tab:surveys}). 
Surveys at low frequencies will likely be sensitive to a different pulsar population, and therefore an all-sky survey at low frequencies is also essential to develop a comprehensive picture of neutron-star/pulsar populations in the Galaxy. Bearing this in mind (and as we detail in \S~\ref{sec:surveysensitivity}), the survey is designed to  reach a final sensitivity comparable to that of LOTAAS, i.e., the use of long dwell times (4800\,s) to attain a limiting sensitivity (10$\sigma$) of $\sim$2-3\,mJy for long-period pulsars with small duty cycles, and assuming a spectral index $\alpha = -1.5$ and no turnover down to $\sim$150\,MHz. This is $\sim$3-5 times deeper than the previous-generation low-frequency (70\,cm) survey \citep{70cm} in the south (and thence an accessible search volume $\sim$5-10 times larger), and $\sim$2-3 times deeper than the high-latitude segment of the Parkes HTRU survey \citep{keith2010}.

The SMART survey will also serve as a reference survey for future deeper surveys at low frequencies, such as those planned with SKA-Low \citep{keane2015}. While the success of (and the lessons learned from) all ongoing low-frequency surveys will indeed inform SKA-Low pulsar surveys, the SMART survey will potentially play an additional important role, since the MWA is also the official low-frequency precursor for SKA-Low, and is located at the same site where SKA-Low will be built. Specifically, the sky coverage of the SMART survey is identical to that of SKA-Low, which means a higher degree of synergistic overlap in calibration and beamforming methodologies, than most northern facilities. The role of reference surveys is vividly demonstrated by the later generation multibeam surveys in the south; e.g., the PMB survey for its successors, the HTRU pulsar survey \citep{keith2010} and the SUrvey for Pulsars and Extragalactic Radio Bursts (SUPERB) \citep{keane2018}, which can now play a similar role for the planned surveys with MeerKAT. However, aside from the Parkes 70\,cm survey of the 1990s, the low-frequency southern-sky remains essentially unexplored for pulsar searches, especially at $\lesssim$300 MHz.

Aside from the aforementioned primary science goals, there are also 
some auxiliary  goals for the SMART survey, largely enabled by the novelty of the data recording strategy, i.e., the use of voltage capture system and post-processing, as opposed to the beamformed data in the filterbank format.
 These not only facilitate a number of additional strategies for confirmation and follow-up, but they can also be potentially exploited for developing and trialling alternate strategies for pulsar searches; 
e.g., image-based techniques for the identification of promising candidates that take advantage of pulsar properties such as steep-spectrum, variability or circular polarisation \citep[e.g.,][]{sett2022}. 
These, in principle, also offer some advantages over traditional search methods, especially for extreme pulsars like those with sub-millisecond periods, or distant pulsars whose pulse shapes will be significantly broadened due to multi-path scattering, but will be  sensitive primarily to very bright sources.

Notwithstanding the anticipated scientific merits of the SMART survey, computational requirements are substantial, especially given the large data rate of the VCS and searching at low frequencies, thereby necessitating a multi-pass processing strategy. In the first-pass processing, we perform a shallow survey, where 10 minutes of data from each observation are processed, and the search is limited to basic periodicity, and DMs up to 250\,\dmu. 
In this paper, we outline the observing strategies employed for the survey, and processing strategies adopted for the initial phase, and present analysis and results to date, as well as plans and strategies for future processing. A companion paper (hereafter Paper II) will describe the survey status, pulsar census to date and more details on follow-up strategies including timing and imaging follow-ups.

\subsection{Survey strategy}
\label{sec:surveystrategy}

The novel strategy employed for the SMART survey, i.e., the use of VCS recording from 128 tiles, which allows high-time resolution (and instantaneous) sampling of a very large patch of the sky (but at the expense of a large data rate of 28\,\TBhr), necessitates substantial processing to enable large-scale pulsar searching applications. Most importantly, the voltage data from the tiles need to be coherently combined to generate thousands of tied-array beams prior to any search processing. The undertaking of the SMART survey is particularly enabled by the Phase II upgrade, whereby a compact configuration of 128 tiles within $\sim$300\,m became available on a semi-regular basis. The compact configuration of Phase II brings an enormous efficiency in terms of beamforming cost; 
specifically, the number of tied-array (i.e. phased array) beams required to fill the full FoV (at a gain level down to half power point) is reduced from $2.7 \times 10^5$  for the Phase I array to $3.9 \times 10^3$  for the Phase II compact array. This reduction of more than two orders of magnitude in the computational cost makes an all-sky high-sensitivity pulsar search tractable (and affordable) with an interferometric array like the MWA. Thus, with the beamforming step integrated into software-defined instrumentation, this effectively translates into an impressively large survey speed of $\sim$450\,\sqdegphr, i.e., the full visible sky of the MWA ($\delta < +30\deg$) can be surveyed in a modest number of VCS pointings.

The first-pass survey strategy of processing only 10 minutes of data from each observation (hence reaching about one-third of the full-search sensitivity) was adopted also to boost the prospects of early pulsar discoveries. Even though the combination of the VCS mode and the FoV provides a large survey speed, practical considerations such as the availability of the compact configuration necessitated multiple  observing campaigns to advance the survey. Further details including 
the survey status and completion plans are described in Paper II.

\subsection{Beamforming and Sky tessellation }
\label{sec:beamforming}

The signal processing chain of the MWA including the high time resolution system is described in a number of earlier papers \citep[e.g.,][]{tingay2013,prabu2015,tremblay2015}, and is briefly reiterated here. In the legacy system that was employed for survey campaigns to date, the VCS sub-system follows the second stage of channelisation in the signal path. Each element of the array is a $4 \times 4$ dipole array, called a ``tile'', the signals from which are fed to an analogue beamformer that defines the FoV. The beamformed signals are Nyquist-sampled at 655.56\,Msps and channelised (after signal conditioning) using a polyphase filterbank (PFB) to generate 256$\times$1.28-MHz signal outputs (i.e., coarse channelisation), 24 of which are transported to the central processing facility, where a second-stage PFB operation is performed, resulting in 128$\times$10-kHz time series for each coarse channel, i.e., 3072 channels across the recording 30.72\,MHz bandwidth. These voltage time series are written to an array of RAID disks by the VCS as 4+4-bit complex voltage samples. 
These data are recorded (up to a maximum duration of 100 minutes) and transported to the Pawsey Supercomputing Centre where further processing (including calibration and beamforming) is carried out. 

VCS-recorded data can be processed offline for calibration and tied-array beamforming \citep{ord2019} and, optionally, can also be reprocessed to reconstruct a higher time resolution voltage data at the native coarse channel resolution of 0.78\,\us{} \citep{mcSweeney2020}. To realise the SMART pulsar survey, this beamformer functionality was further enhanced 
to optionally generate several dozens of tied-array beamformed outputs simultaneously -- i.e. the so-called {\emph{multi-pixel}} beamformer, which is essentially the front-end of the pulsar search processing chain. The implementation details and benchmarks are described in \citet{swainston2022}.
This software tied-array beamformer has been benchmarked on Pawsey's Garrawarla and Swinburne's OzSTAR supercomputers. It performs $3\times$ faster on the latter, which has been the primary high performance computing (HPC) platform for much of our SMART data processing. 

Thanks to the large field-of-view of the MWA ($\sim$610\,\sqdeg at 155\,MHz, near zenith), the entire sky south of declination $\delta < +30\deg$ can be covered in a modest number of telescope pointings. The sky tiling strategy is shown in Fig.~\ref{fig:skytessellation}. In short, we adopted pointings similar to that of the GaLactic and Extragalactic All-sky MWA  survey \citep{wayth2015}, i.e., meridian drift scans optimised for  maximum sensitivity at each declination as well as for more reliable calibration (referred to as `sweet spots'). In this case, the number of pointings depends on the degree of overlap in right ascension (RA), with a minimum of 58 pointings for minimal ($1\deg$) overlap and 78 for a $15\deg$ overlap. A large overlap is more optimal as it effectively serves as a two-pass strategy, which is desirable at low frequencies where intermittency (from effects such as scintillation) tends to be more pronounced. 
After exploring the full range of options, and also factoring in the available resource constraints,
we converged on a $10\deg$ overlap as an acceptable choice.\footnote{The operational constraints of the MWA limited 
VCS mode observations to a maximum of 25 hours per observing semester,
with the legacy system.}
As shown in Fig.~\ref{fig:skytessellation}, this amounts to a total 70 pointings, i.e., 93\,hr of telescope time for the full SMART survey.

For each pointing, many tied-array beams (TABs) are formed to maximise the sensitivity across the FoV. The tied-array beams are pointed towards fixed right ascension and declination, with the necessary adjustments to the tile phases made for every second of data \citep{ord2019}. Thus, although the observations themselves are drift scans, sources can be tracked by the same TAB for up to the full duration of the observation.

The precise size and shape of the TABs is a non-trivial function of the tile layout of the compact configuration, equivalent to the ``Compact robust 1’’ synthesised beam whose cross section is presented in Figure 7 of \citet{wayth2018} and discussed in \citet{swainston2022} and in Section \S\ref{sec:dataprocessingandanalysis}. Due to the compact configuration’s redundant baselines (in the two ``hexes’’), the most sensitive parts of the TAB consist of a main lobe whose full width half maximum (FWHM) at 155\,MHz is 23$^\arcmin$, surrounded by a pattern of discrete grating lobes of similar width. Although these grating lobes can be exploited for candidate confirmation (further discussed below), we choose the TAB pointings to form a dense (hexagonal) grid such that the main lobes overlap by $\sim$20\%, as shown in Fig.~\ref{fig:skytessellation}. This effectively Nyquist-samples the sky at a gain of the half-power level or more. The beam shape used for this calculation assumes that all 128 tiles are functioning, whereas, in reality, up to $\sim$10\% of tiles may be flagged in any given observation. Unless the flagged tiles preferentially result in a reduction of the longest baselines,  the effect on the beam shape is negligible.

Tiling the FoV in this way translates to $\sim$6300 TABs for an observation pointed toward the zenith. For pointings away from the zenith, where the beam shape develops a significant ellipticity (e.g., at zenith angle 15\deg, ellipticity $\epsilon = \theta _{maj} / \theta _{min} =$ 1.36 where $\theta_{maj}$ and $\theta _{min}$ are the major and minor axes of the TAB), the number of TAB pointings are in the range $\sim$4200-4500. 
Further, the beam size also varies across the 20\% fractional bandwidth of our survey observations; for example, for a pointing toward the zenith (where the TAB is nearly symmetrical), the FWHM is 25.3$^\arcmin$ at 140\,MHz but reduces to 20.7$^\arcmin$ at 171\,MHz. This further justifies our rationale for a 20\% overlap, as it ensures every single spot in the sky is covered at a gain near or above the half power level even at the high end of the observing band. 

Finally, as with any other aperture array, the sensitivity is not uniform across the sky and is strongly declination dependent; to first order, the loss in sensitivity is by a factor $\cos(\theta _z)$ where $\theta _z$ is the zenith angle. In principle, this can be compensated to a certain extent by longer integrations, though
in practice, the inherent limitations of our data recording system (VCS) limits this to no more than 90 minutes, and we therefore use 80 minute recordings for all pointings. As such, the sensitivity will not be uniform across the sky due other factors; e.g., the sky background temperature \Tsky{} is direction dependent, and the loss in sensitivity from severe pulse broadening for distant pulsars, which applies to the sight lines within the Galactic plane or toward the Galactic centre. Some of these are considered in detail in \S~\ref{sec:surveysensitivity}.

\begin{figure*}[t]
\begin{center}
\includegraphics[width=0.48\linewidth]{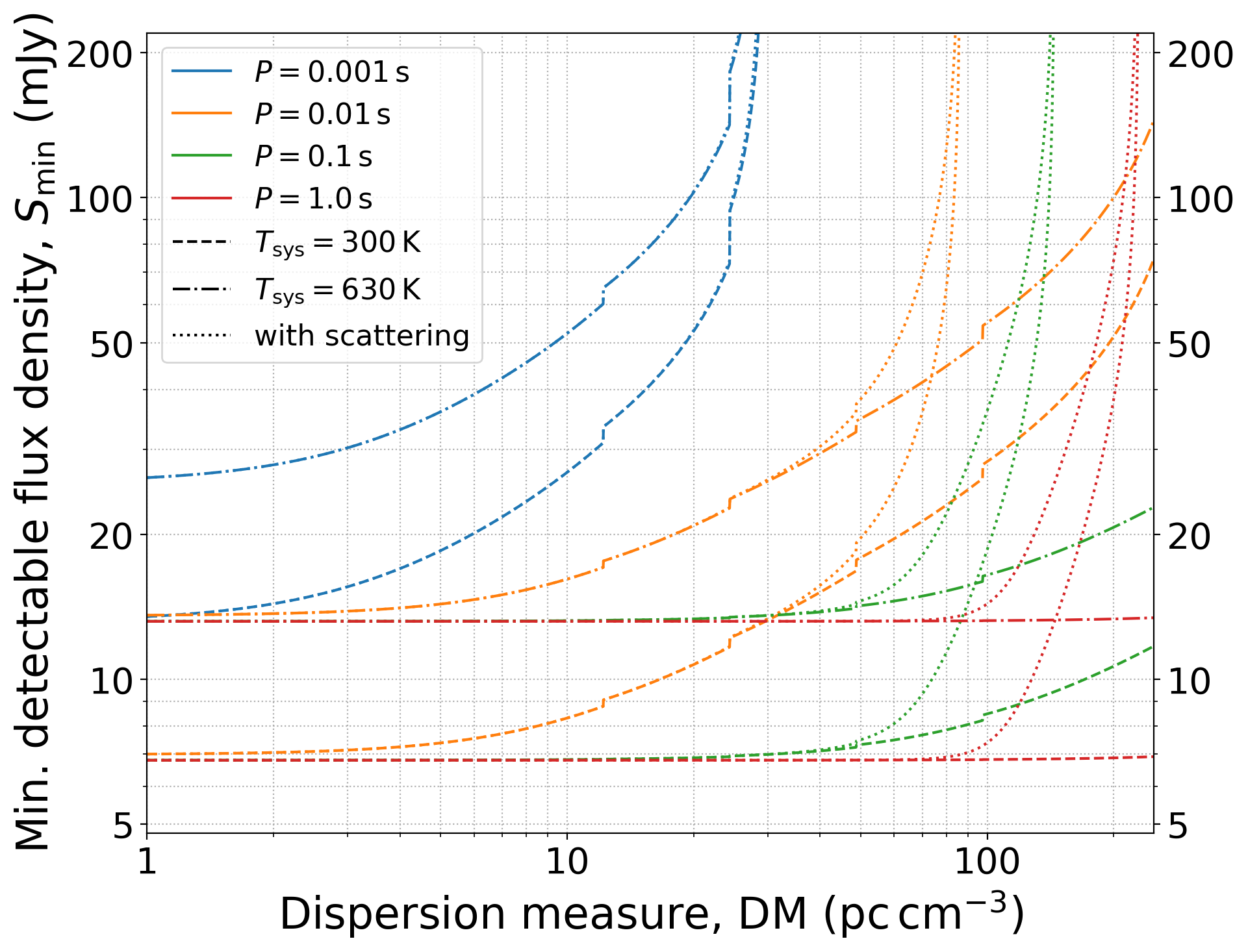}
\includegraphics[width=0.472\linewidth]{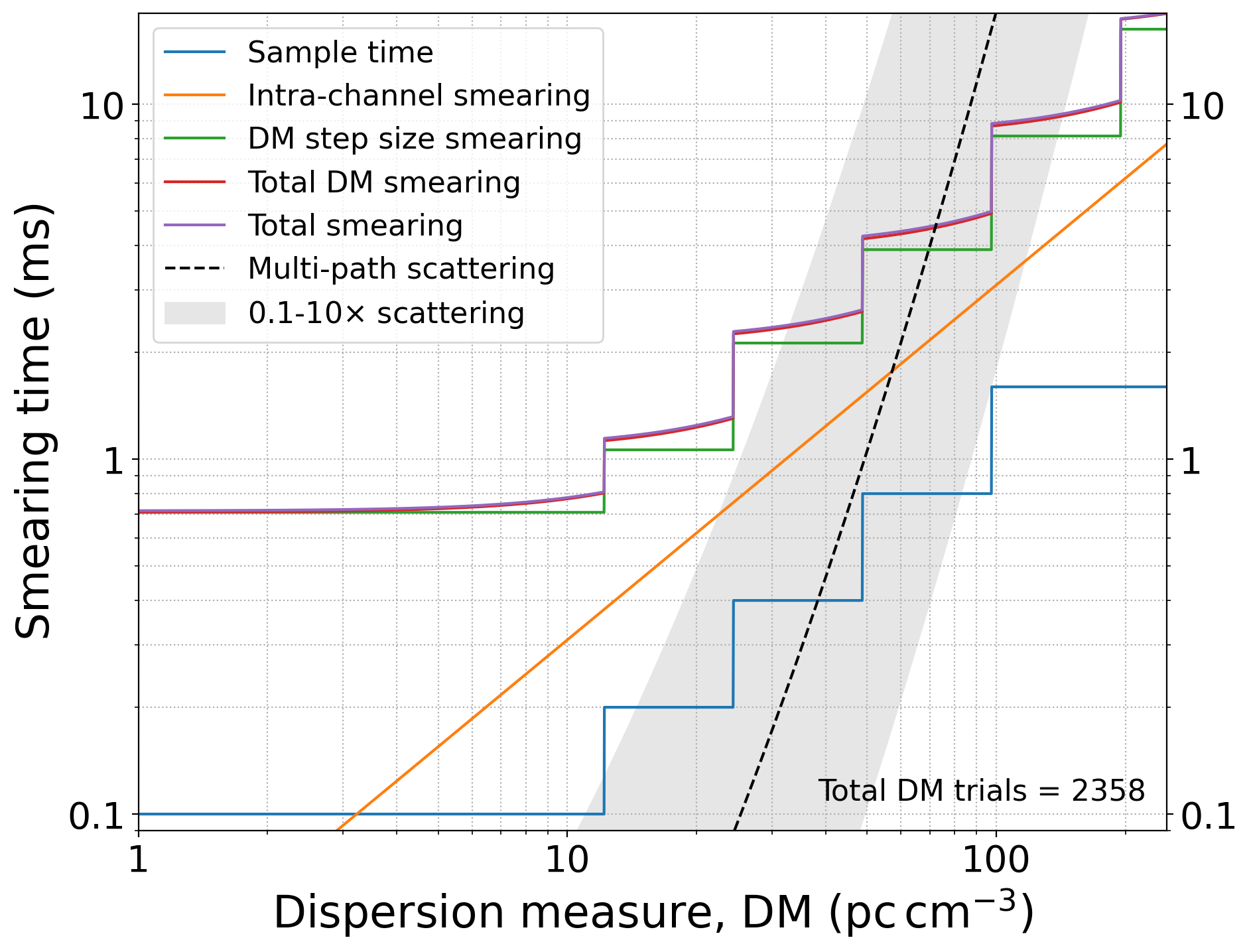}
\caption{
\textit{Left}: Minimum detectable flux density, \Smin, for the first-pass processing of the SMART survey as a function of DM. Sensitivity limits, assuming a 10-minute integration time, are plotted for different pulse periods, $P=$ 1.0, 0.1, 0.01, 0.001\,s, and for two different system temperature values \Tsys; one corresponding to mean \Tsky{} for regions away from the Galactic plane, and the other for a mean \Tsky{} in the plane, but excluding the region toward the Galactic Centre. The effect of pulse broadening due to interstellar scattering (Bhat at el. 2004) is shown by the dotted lines. \textit{Right}: Pulse broadening (smearing) incurred by using the first-pass processing dedispersion plan (Table~\ref{tab:ddplanfirst}) due to various factors such as the finite sampling time, dispersive smearing due to the incoherent de-dispersion algorithm used, and the effects of multi-path scattering based on the $\tau_d$-DM relation from Bhat et al. (2004). The grey shaded region denotes one order-of-magnitude larger or smaller range in the predicted scattering. 
}
\label{fig:surveysens}
\end{center}
\end{figure*}

\subsection{Survey sensitivity }
\label{sec:surveysensitivity} 

The sensitivity of a pulsar survey is determined by the combination of some instrumental and processing parameters and a variety of broadening effects to pulsar signals. Following \citet{dewey1985}, the minimum detectable flux density for a pulsar with period $P$ and effective pulse width \Weff, down to detection significance $({\rm S/N}) _{\rm min} $, i.e., minimum detectable signal-to-noise ratio, is related to the telescope gain $G$ and system temperature \Tsys, which is the sum of the receiver and sky background temperatures, i.e., 
\begin{equation}
    \Smin =  \frac{ ({\rm S/N})_{\rm min} ( \Trecv + \Tsky )}{ G \sqrt{ \Npol \Tobs \Bobs } } \sqrt { \frac{ \Weff}{ P - \Weff } } 
\end{equation}
where \Npol{} is the number of polarisations summed, \Tobs{} is the integration time and \Bobs{} is the recording bandwidth. As evident from this equation,
the sensitivity is maximum for long-period pulsars with a small duty cycle,
i.e. when $ \Weff \ll P  $. The gain $ G = \Aeff  / 2 k_B $, where \Aeff{} is the effective collecting area and $k_B$ is the Boltzmann constant. 
At 150 MHz, 
$ \Aeff  \approx 2750 $ \sqm for a 128-tile MWA \citep{tingay2013}, 
which may imply $ G \sim 1 $\,\KperJy, however, for an aperture array such as the MWA, it is a strong function of the zenith angle, i.e. 
$G(\theta _z) = G_{\rm max}  {\rm cos}(\theta _z)$, where $\theta _z$ is the zenith angle and $G_{\rm max}$ is the gain at $\theta_z=0$. Moreover, for drift-scan type observations that we employ for the SMART, $G$ depends on the offset from the phase centre, and can be $\sim$0.5\,$G_{\rm max}$ at the half power point. We therefore assume a conservative $G \sim 0.5$\,\KperJy for all our sensitivity calculations. 
This is assuming a full coherent beam sensitivity, i.e., perfect calibration for TAB formation and no loss of sensitivity due to flagged tiles. In practice, a small number of tiles ($\lesssim$10) are typically flagged due to malfunctioning, sub-optimal performance or poor calibration solution. As we detail in \S\ref{sec:dataprocessingandanalysis}, the strategy of observing multiple calibrators for SMART observation allows us to perform useful cross-checks and maximise the achievable sensitivity using the best available calibration solutions. 

At the low frequencies of the MWA, the system temperature \Tsys{} is dominated by the sky background \Tsky. Both \Trecv{} and \Tsky{} are frequency dependent, 
and  \Tsky{}  is also a strong function of the direction $ (l,b) $, where $l$ and $b$ are the Galactic longitude and latitude, respectively. We assume a mean \Trecv = 50\,K for the 140-170\,MHz band. Excluding a $\sim$10$\deg$ cone around the Galactic centre, \Tsky{} can vary from $\sim$200\,K toward $|b| \gtrsim 60\deg$  to as much as $\sim$1200\,K in the plane, toward $\gtrsim$10$\deg$ from the Galactic centre, where \Tsky{} can be as large as $\sim 10^4 $\,K at 155\,MHz. We use the \citet{haslam1982} map as the reference and assume $ \Tsky \propto \nu^{-2.55} $ scaling from \citet{lawson1987}. Given this strong dependence of \Tsky{} with $(l,b)$, we consider two cases: (1) the sky at $ |b| \lesssim 5\deg$ where mean $\Tsky\sim 600$\,K and (2) the sky at $ |b| \gtrsim 5\deg $, where mean $\Tsky\sim 270$\,K; i.e., $\Tsys = 630$\,K and $300$\,K, respectively, as shown in Fig.~\ref{fig:surveysens}.

Intrinsic pulses are broadened due to a variety of effects, as discussed earlier.
As detailed in \citet{handbook2012}, the total smearing time \tautot{} is the quadratic sum of the finite sampling time \tausamp, the residual dispersive smearing due to finite frequency channel \tauchan, the dispersive smearing across the full recording band due to finite DM steps in trial DM values \taudm, and the dispersive smearing resulting from piece-wise linear approximation of the quadratic dispersion law in the sub-band dedispersion algorithm employed in searches \tausub. Fig.~\ref{fig:surveysens} summarises these for our current first-pass processing.
The planned second-pass search will significantly enhance the search sensitivity by processing the full observation (4800 s) and the use of more optimal DM steps, i.e. many more trial DMs than that used in current search.

As evident from the figure, for our current first-pass processing, the total smearing time is dominated by finite DM steps; this sub-optimal choice was made in an effort to maximise the number of observations that can be processed to completion toward a shallow all-sky survey within available computing resources. The dedispersion plan 
utility used is shown in Table~\ref{tab:ddplanfirst}. In effect, we progressively downsample the data five times over the DM range searched, each time making the DM step size coarser. At ${\rm DM} \gtrsim 3$\,\dmu, the dispersive smearing time within the 10-kHz channel is larger than the native sampling time (100\,$\mu$s) but still a smaller contribution to the total smearing time, compared to that due to the DM step size. As a result, the net smearing time \tautot\ displays a step-wise increase as shown in Fig.~\ref{fig:surveysens}, given our dedispersion plan.  
At very low DMs $\lesssim$10\,\dmu, $\tautot \sim 0.7$\,ms but increases to $\sim$10\,ms at DM $\sim$100\,\dmu. In essence, our first-pass search severely compromises the sensitivity to millisecond pulsars (MSPs) at larger DMs and shorter periods, i.e., 
it is currently sensitive to MSPs at $\text{DM}\lesssim$30\,\dmu and $P \gtrsim$10\,ms. As shown in Fig.~\ref{fig:surveysens}, at those larger DMs, the smearing due to scattering (i.e. pulse broadening) can also be significant. The broadening time here is based on the empirical relation in \citet{bhat2004}, which is mostly relevant for pulsars near the plane. As is well known, these scattering estimates can be uncertain by more than an order of magnitude, denoted by the grey shaded region.

\begin{table}
\begin{threeparttable}
\caption{Dedispersion plan for the first-pass SMART processing} \label{tab:ddplanfirst}
\begin{tabular}{cccccc}
\hline
  ${\rm DM} _{\rm min}$ &
  ${\rm DM} _{\rm max}$ &
  $\delta {\rm DM}$ &
  $N _{\rm DM}$ &
  $d_s$ &
  $\Delta t_{\rm eff}$ 
\\
  \multicolumn{1}{c}{$ (\dmu) $} &
  \multicolumn{1}{c}{$ (\dmu) $} &
  \multicolumn{1}{c}{$ (\dmu) $} &
  \multicolumn{1}{c}{} &
  \multicolumn{1}{c}{} &
  \multicolumn{1}{c}{(ms)}
\\
\hline   
    1.0 &    12.2 &    0.02 &    560 &        1 &       0.1\\
   12.2 &    24.4 &    0.03 &    406 &        2 &       0.2\\
   24.4 &    48.8 &    0.06 &    406 &        4 &       0.4\\
   48.8 &    97.6 &    0.11 &    443 &        8 &       0.8\\
   97.6 &   195.2 &    0.23 &    424 &       16 &       1.6\\
  195.2 &   250.0 &    0.46 &    119 &       16 &       3.2\\
\hline
\end{tabular}
\begin{tablenotes}
\item The columns 1 and 2 denote the ranges in dispersion measure, between ${\rm DM} _{\rm min}$ and ${\rm DM} _{\rm max}$, with a DM step-size of $\delta {\rm DM}$, resulting in $N _{\rm DM}$ trial DM values. The down sampling factor is denoted by $d_s$, i.e. the factor which the temporal resolution is averaged to yield a net resolution $\Delta t _{\rm eff}$. 
\end{tablenotes} 
\end{threeparttable}
\end{table}

The theoretical sensitivity is shown in Fig.~\ref{fig:surveysens} for different periods, $P=1.0$, $0.1$, $0.01$ and $0.001$\,s. 
In all these calculations, we have assumed a duty cycle of 3\%, i.e., $W_{\rm eff}/P=0.03$. 
For each period, a pair of curves are shown: one for the best-case scenario, i.e., searches away from the plane, where $\Tsys\sim 305$\,K; and the second for the sky near the plane where the mean \Tsys{} is twice as high. 
In either case, the sensitivity is maximum for long-period pulsars, at low to moderate DMs of $\lesssim$50\,\dmu, and toward $|b| \gtrsim 5\deg$. 

With our first-pass processing scheme (i.e., 10-minute integrations and a sub-optimal dedispersion plan), we reach a limiting sensitivity of $\Smin\sim 7\text{-}12\,\mJy$ for long-period pulsars, and $\sim 12\text{-}25\,\mJy$ for MSPs at low to moderate DMs.
For the proposed deep-pass processing (i.e., $\sim$80-minute integrations and a more granular dedispersion plan), we can achieve
a limiting sensitivity of $\Smin \sim 2\text{-}3\,\mJy$ for long-period pulsars and $\sim 5\text{-}10\,\mJy$ for MSPs at low to moderate DMs.
In this case, 
the SMART survey sensitivity is comparable to that of the LOTAAS survey in the northern hemisphere. 
While LOTAAS can be twice as sensitive as SMART for long-period pulsars, the sensitivity for $P \lesssim$10\,ms is almost similar, owing to a lower degradation in sensitivity in the SMART band. 
Compared to the Southern Pulsar Survey of the 1990s at 430 MHz (i.e., a wavelength of 70 cm), also known as the Parkes 70cm survey \citep{70cm}, the SMART survey is $\sim$3-5 times more sensitive, especially for pulsars at DMs $\lesssim$100\,\dmu and spectral index $\alpha \lesssim -1.5$. 
Even the ongoing shallow survey is comparable to the 70cm survey in theoretical sensitivity, and if at all, slightly more sensitive for steep spectrum pulsars with no turnover down to $\sim$100\,MHz. 
This provides a strong motivation to undertake a full-scale pulsar survey with the MWA.

\subsection{Effective dwell time and sensitivity}
\label{sec:effective_dwell_sens}
Unlike most other pulsar surveys, where single-dish telescopes are used to track targeted positions for small time intervals (e.g., HTRU, GBNCC), the SMART observations are drift scans, where the primary beam is pointable but static in horizontal coordinates (azimuth and zenith angle) once an observation starts and the sky moves through the FoV. 
When forming TABs, we track the sky position as it moves through the MWA primary beam and as a consequence not all TABs necessarily remain within a sensitive part of the primary beam for the full 80-minute duration. 

\begin{figure}[tb]
    \centering
    \includegraphics[width=\textwidth]{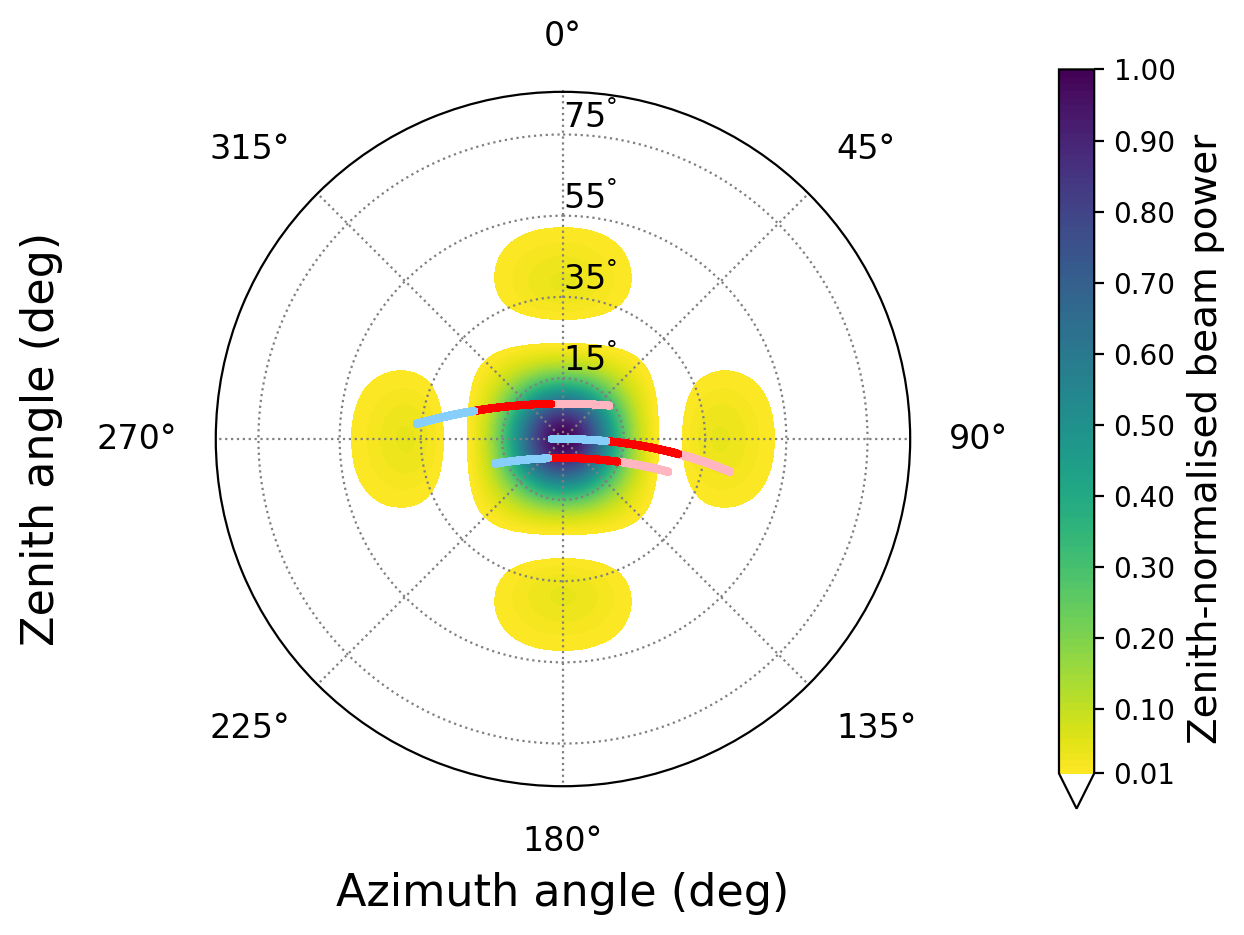}
    \includegraphics[width=\textwidth]{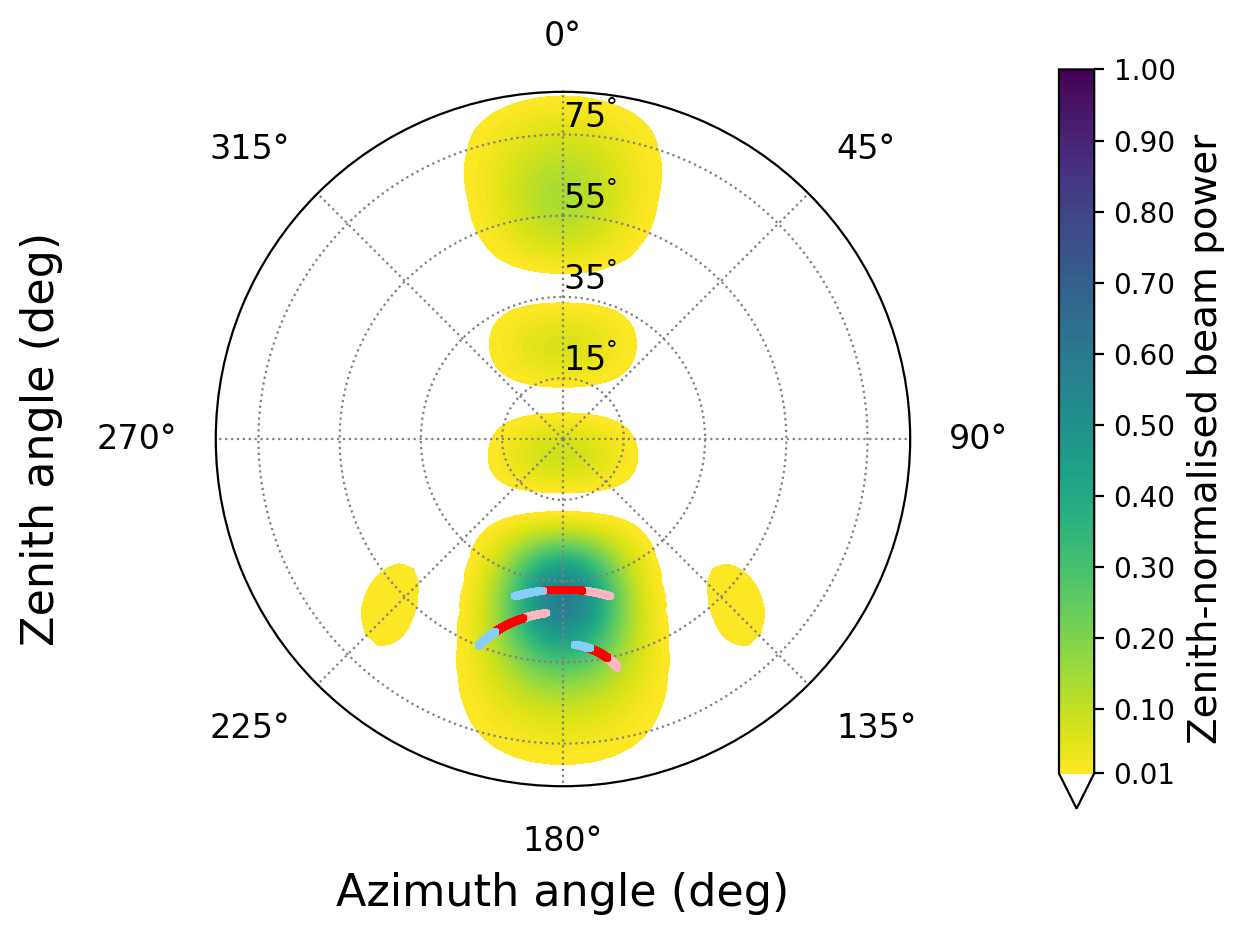}
    \caption{Tied-array beam traces through the MWA primary beam for SMART observations. Three example pointing directions for each observation are traced including 1\,hr before and 1\,hr after the 80-minute observation. The target trace (rotating clockwise as time advances) is coloured pink to represent the trajectory before the observation, red during the observation, and blue after the observation is complete. North is at $0\deg$ and the azimuth angle increases to the East. The colour scales are the same for each subplot, highlighting the sensitivity penalty incurred for observing away from zenith.}
    \label{fig:tab_tracks}
\end{figure}

The amount of time spent within an individual observation FWHM depends both on the observing declination (i.e., where the primary beam is pointed) and the target source position to be tracked with a TAB.
As an example, in Fig.~\ref{fig:tab_tracks} we plot some representative TAB pointings along with the primary beam response for the same observations as in Fig.~\ref{fig:skytessellation} in horizontal coordinates. 
As already noted, our sensitivity drops substantially as we observe at larger zenith angles, which we visualise by having the colour scale represent the zenith-normalised primary beam power as a proxy for sensitivity.
Secondly, the TAB pointing directions are traced before, during, and after the 80-minute observation, which highlights that not all targeted positions remain in a usable part of the primary beam.
These effects highlight at least three points for consideration: (1) it will be an inefficient use of resources to track certain pointings for the full observing duration, (2) tracking pointings naively for the full duration, especially if a significant fraction of the time is spent below the 10\% power point, may actually reduce sensitivity to pulsars, and (3) the full-sky sensitivity will be patchy regardless of TAB forming strategies (although this is partly mitigated by having observations overlap by $\sim$20\% at the central frequency). To address (1) we can estimate the time a source remains in a reasonable power range of the primary beam and only form TABs from the appropriate subset of voltages recorded (e.g., while the target source is not in a null). For (2) we must strike a careful balance between achieving maximal sensitivity (by cutting off parts of the TAB) and dwell time (which benefits searches for longer-period pulsars and single pulse events). The consequence of (3) is unavoidable given the telescope configuration and observing strategy employed, but is quantifiable. 

\begin{figure}[tb]
    \centering
    \includegraphics[width=\textwidth]{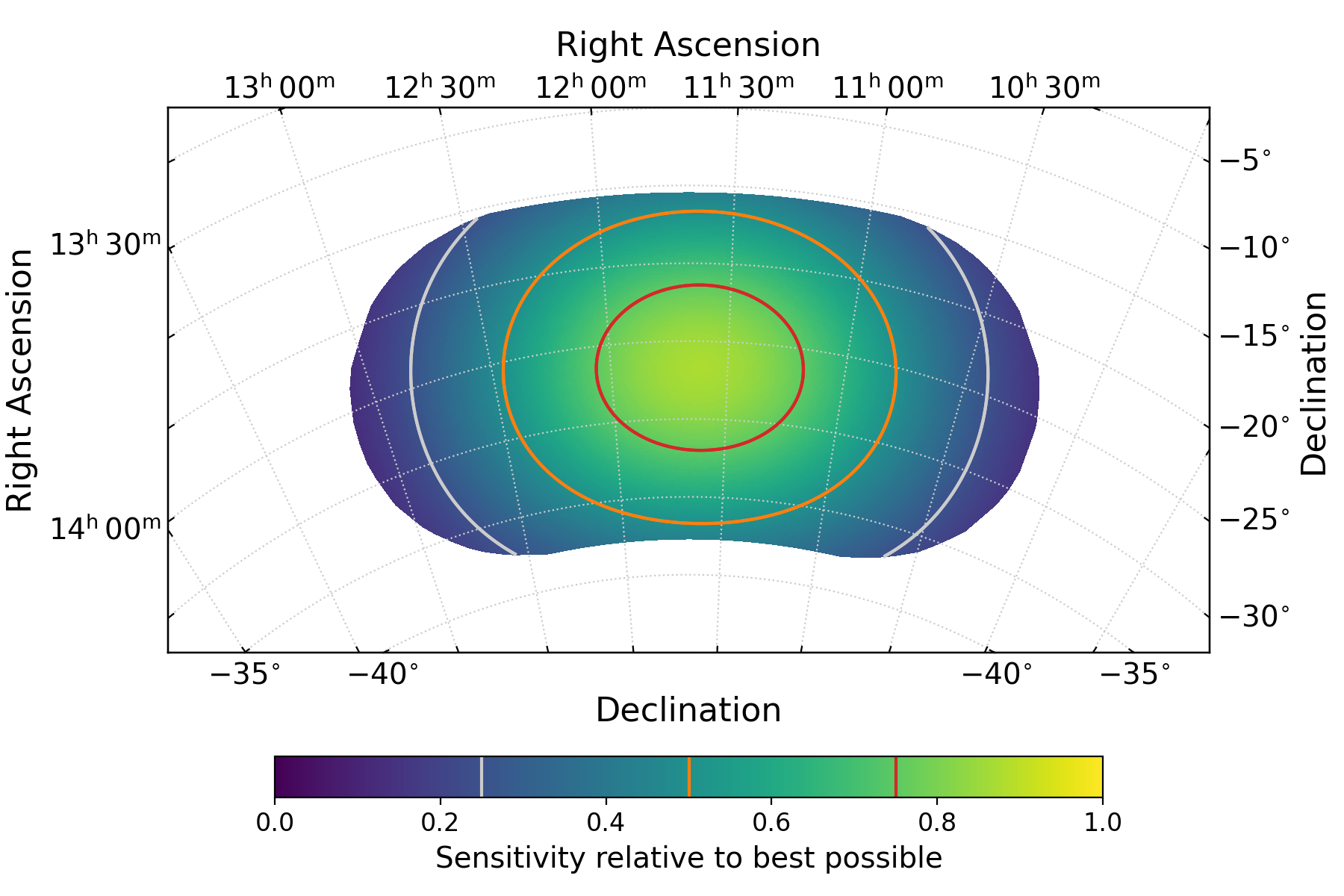}
    \includegraphics[width=\textwidth]{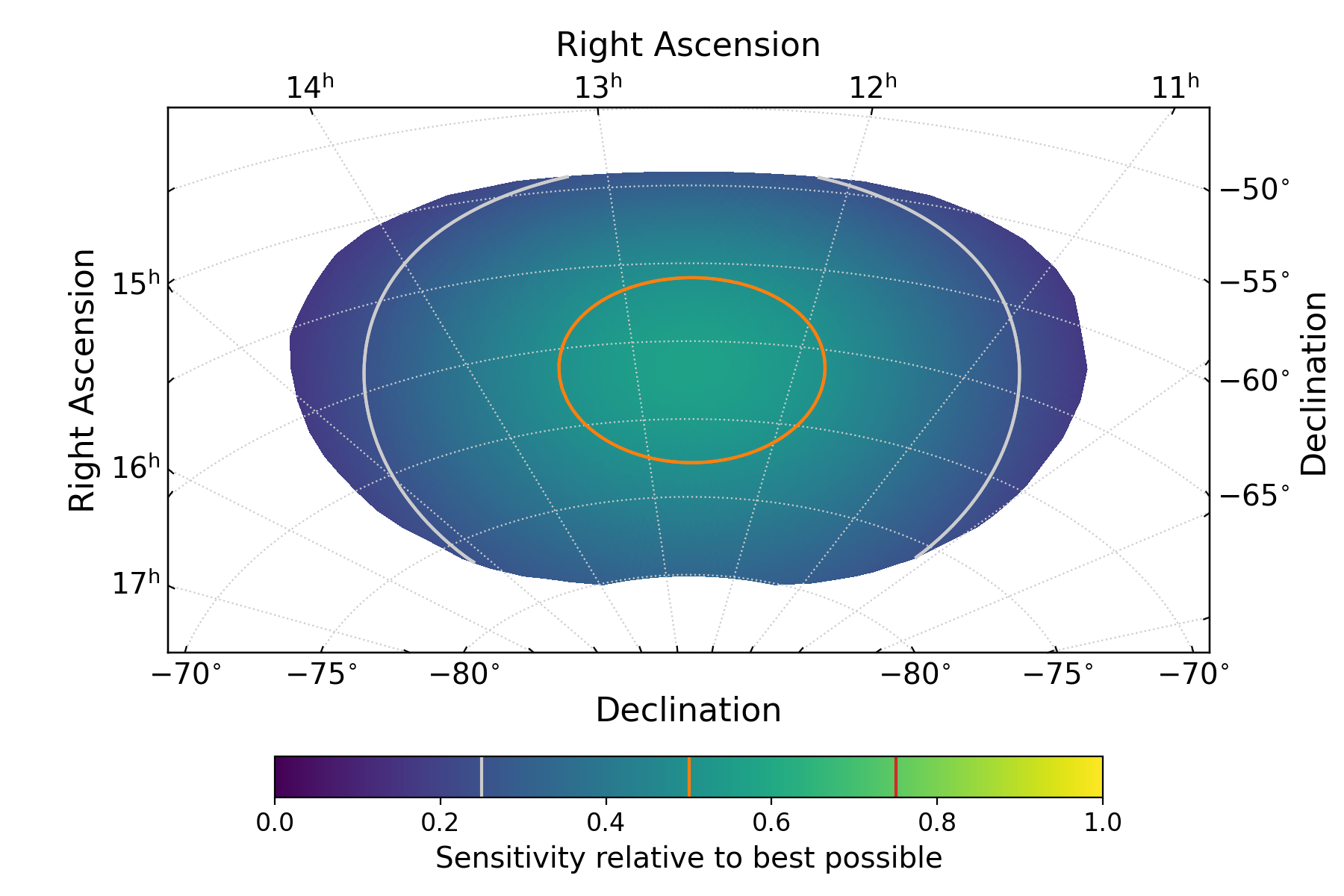}
    \caption{Effective sensitivity maps, assuming a full 80-minute tracking and integration for a given TAB sky position. The colour map is normalised to the best possible sensitivity (described in the text), and contours at 25, 50, and 75 per cent are drawn for clarity. Due to the drift scan nature of the observations versus the tracking TABs, we can never achieve the best possible sensitivity. Right Ascension and Declination are marked by the vertical and horizontal curved grid lines, respectively.}
    \label{fig:effective_sensitivity}
\end{figure}

We can evaluate the relative sensitivity (assuming a 80-minute track for a given TAB) by summing the primary beam response power at discrete time steps, where we use our current best Full Embedded Element (FEE) model \citep{fee_beam}, normalised to the equivalently summed power that would represent the best possible dwell time and sensitivity combination. 
For our purposes we define this quantity as the sum of the primary beam power at zenith for the full observing duration (i.e., imagining we can track an equatorial position with full zenith sensitivity).
This is useful as it scales the effective sensitivity to a quantity close to what a single-dish steerable telescope could achieve. 
In Fig.~\ref{fig:effective_sensitivity} we present these effective sensitivity maps, in equatorial coordinates, for the same example observations used in Fig.~\ref{fig:tab_tracks}.


\begin{figure*}
\centering 
\includegraphics[width=0.64\textwidth,angle=0]{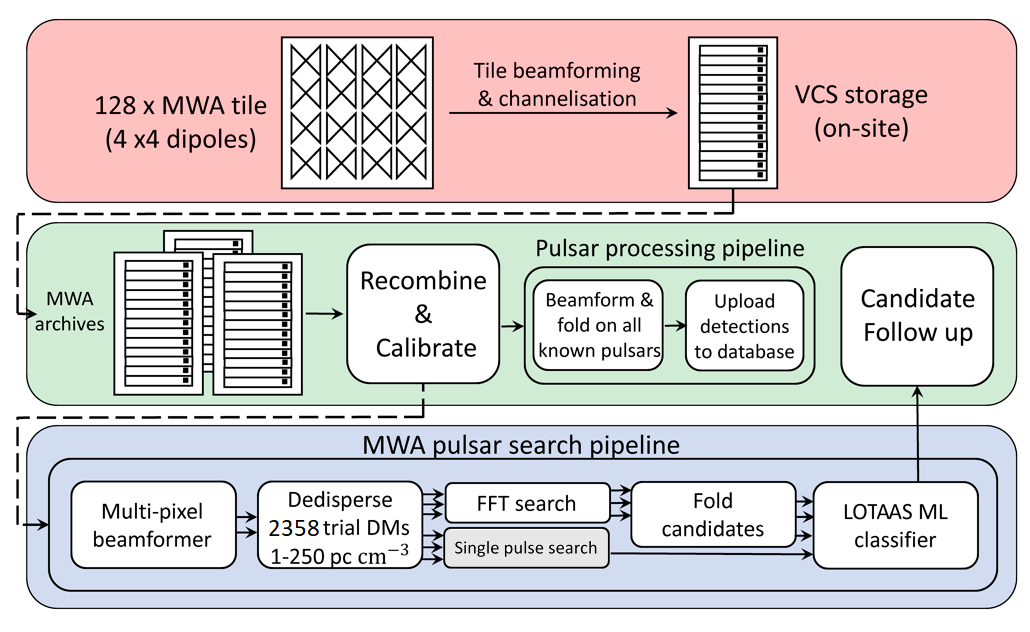}
\caption{
Workflow diagram illustrating the first-pass SMART processing pipeline: voltage data at 100-$\mu$s/10-kHz resolutions are recorded from 128 tiles of the array after tile beamforming and channelisation stages,
and are subsequently ported to the Pawsey supercomputer where the initial processing including calibration, beamforming and known pulsar detections are carried out. Search processing is currently performed on the OzSTAR supercomputer, and is 
limited to basic periodicity searches. 
}
\label{fig:smartpipeline}
\end{figure*}

\section{Data processing and analysis }
\label{sec:dataprocessingandanalysis}


In terms of data collection and processing requirements, the SMART survey is the largest all-sky pulsar survey undertaken in the southern hemisphere, and is only the second largest after LOTAAS. 
The SMART survey will accrue $\sim$3\,PB of VCS data, compared to $\sim$1\,PB (search mode data) by the highly successful Parkes HTRU survey, and $\sim$8\,PB (beamformed data) by LOTAAS. 
As outlined earlier, the survey will cover the sky in 70 VCS pointings, each VCS observation being 4800\,s (42\,TB). The management and processing of this volume of data is non-trivial, particularly considering the computational resources currently available. The processing software and pipelines are developed, tested and benchmarked on Pawsey's Galaxy/Garrawarla clusters, and subsequently ported and benchmarked on Swinburne's OzSTAR supercomputer.
The time on OzSTAR is secured via the merit allocation scheme under Astronomy and Supercomputer time allocation, and is typically 0.5-0.6 million service units (CPU core) hours per annum. These constraints largely drive the initial processing strategies, thereby necessitating a first-pass shallow survey.

Compared to the HPC resources available at Pawsey, the processing efficiency has been relatively higher on OzSTAR, where the current benchmarks are 2 kSU for beamforming and 25 kSU for searching a 10-min observation (4.4\,TB), where 1 kSU = 1000 service units (CPU core hours). The current allocation thus allows processing of 9 observations (fields) per semester, where each 10-min VCS observation is processed for $\sim$6000 tied-array beams, each of which is then searched in 2358 trial DMs, out to 250\,\dmu. The completion of first-pass processing will thus require $\sim$2 million core hours. Scaling from the current benchmarks, we would thus expect 1,500 kSU per full observation for deeper searches, and 60 million core hours for full DM searches ($\sim$10,000 searches, for a max DM of 250\,\dmu), necessitating the integration of GPU-based search processing in the future.

An overview of the processing pipeline is presented in Fig.~\ref{fig:smartpipeline}, the details of which are described in the sections below. In essence, this involves preprocessing and beamforming of voltage data from 128 tiles of the array to generate beamformed time series, 
before the data can be processed through the search and detection pipelines. The main steps are outlined below. 

\subsection{Pre-processing and Beamforming}
\label{sec:preprocessing}

The main step in the preprocessing stage involves processing VCS data so they can be calibrated and coherently combined to produce beamformed time series at the native resolution of 100-\us/10-kHz of the VCS. The array calibration is performed using one of the standard calibrators (e.g., 3C444), recorded in the visibility mode at the default 0.5-s/40-kHz resolution, where complex gain solutions (amplitude and phase) are obtained for each of the 128 tiles, for every coarse channel (1.28\,MHz wide), using the Real Time System (RTS) software package. The procedure is essentially similar to those employed for other VCS observations \citep[e.g.,][]{psrone}. The calibration solutions can then be used to coherently combine the voltage data in phase using the tied-array beamformer, the conceptual details and implementation of which are detailed in \citet{ord2019}. The functionality was enhanced, and GPU parallelised, in preparation for SMART data processing \citep{swainston2022}.

The beamformed data are written as Stokes I at 100-\us/10-kHz resolutions. The current implementation allows processing 120 coarse-channel beams at once, i.e., 5 full-bandwidth (30.72\,MHz) beams, resulting in a data rate of 87\,GB/beam for a 10-minute observation. For each survey pointing, this amounts to  $\sim$500\,TB in beam-formed data. These data are equivalent to that would emerge from the standard pulsar backends and so can be processed using standard pulsar search packages. 
Typically, data would be processed to generate RFI masks; however the superb radio-quiet environment at 
the telescope site
and preferential observing during the nightly hours (and within an hour of the source transit) make this step not essential for the SMART data. In most cases, data are minimally affected by RFI, and consequently no RFI-related processing is carried out in the ongoing first-pass processing.

The large FoV of the MWA means excellent prospects for detecting multiple known pulsars within each pointing, which is also important for crucial data quality checks and initial assessment of array calibration and tied-array sensitivity. 
In short, each SMART observation is processed for known pulsars within the primary beam ($\sim$610\,\sqdeg), using a custom pulsar detection pipeline.

\subsection{Search pipeline}
\label{sec:searchpipeline}

The current SMART pipeline includes a GPU-based pipeline for front-end processing (beamforming) and a CPU-based pipeline for downstream (search) processing. The search pipeline is based on the Pulsar Exploration and Search Toolkit\footnote{See \url{https://github.com/scottransom/presto}} \citep[PRESTO;][]{presto, presto_ascl} pulsar search software suite, with the addition of machine-learning (ML) tools adopted from the LOTAAS classifier \citep{tan2018b}. 
This was adopted as a first-pass processing strategy, to ensure an end-to-end working pipeline from the data collection and reordering stage 
(occuring at the observatory site) 
to array calibration/quality checks (Pawsey) and search processing (OzSTAR). 
To encapsulate the full search workflow, we make extensive use of Nextflow\footnote{See \url{https://github.com/nextflow-io/nextflow}} \citep{nextflow} to manage data input, output, processing tasks, and intermediate or final product creation and tracking.

In the near future, as we transition to full-sensitivity searches, the search component will be replaced by a GPU-based implementation.  
Here we present a detailed breakdown of the current SMART search pipeline, where 10-min data (4.8\,TB) are processed from each observation. 

\subsubsection{Dedispersion and periodicity search}
\label{sec:dedispersion}

The beamformed data are processed to create dedispersed time series for each beam. As mentioned earlier, for the first-pass  processing, maximum DM searched is 250\,\dmu. At higher DMs, scattering can be significant; e.g., pulse broadening times $\gtrsim$\,100 ms are expected at 155\,MHz for sight lines toward $|b|\lesssim 5\deg$, and $l\gtrsim 330\deg$ or $l\lesssim30\deg$, where such high DMs can be expected. Further, even with 10-kHz channels, DM smearing can still be significant at low frequencies. For instance, at a frequency of 140\,MHz (i.e., the low end of the SMART band), intra-channel dispersion smearing is $\sim$1.5\,ms at $\rm DM = 50\,\dmu$, and $\sim$10\,ms at $\rm DM \sim 250\,\dmu$. The dedispersion plan was created using the PRESTO {\tt DDplan.py} utility, but with the caveat that sub-optimal settings were chosen (the use of coarser DM steps) to limit the number of DM trials to 2358, given the limitation of computational resources. The {\tt prepsubband} tool from PRESTO was used to create incoherently dedispersed time series from the PSRFITS (i.e., search mode) files. It makes use of the sub-band dedispersion technique, which uses a piece-wise linear approximation to the quadratic dispersion relation. The dedispersion plan employed in the first-pass search is shown in Table \ref{tab:ddplanfirst}.

Searching for periodic signals involves computing the power spectra of the dedispersed time series, which is performed using the {\tt realfft} tool within PRESTO, by applying Fourier transform techniques. These power spectra are then searched for periodicities using {\tt accelsearch} \citep{ransom2002}, which detects the most significant periodic signals and uses harmonic summing to recover the 
power spectra at multiples of a given spin frequency.
No acceleration searches are performed in this first pass; i.e., searches are only performed at zero acceleration. Acceleration searches would require significant processing cost, given the large data rates, and the number of trial DMs required, but will be part of the second pass search.
If the significance of any spectral bin is in excess of $2\sigma$, it is marked as a candidate and the corresponding harmonics up to the 16th are summed to increase the detection significance. 

A sifting procedure is then performed on the list of candidates from all 2358 DM trials. We adopt a fairly standard procedure, quite similar to that followed for LOTAAS, where candidates with $P<1 $\,ms or $P>30$\,s are rejected,\footnote{This period range was adopted given the minimum and maximum period of known pulsars in the ATNF pulsar catalogue when our processing commenced, which was 1.3\,ms and 23.5\,s, respectively.} as well as those with $\rm DM<1$\,\dmu. 
Candidates with similar DMs and harmonically related periods are then grouped, and only the instance with the highest S/N is kept. From this reduced candidate list, only those with $\gtrsim$5$\sigma$ detections are then folded.

\subsubsection{Candidate folding}
\label{sec:folding}

\begin{figure*}[ht]
  \subfloat[]{\includegraphics[width=0.5\textwidth]
  {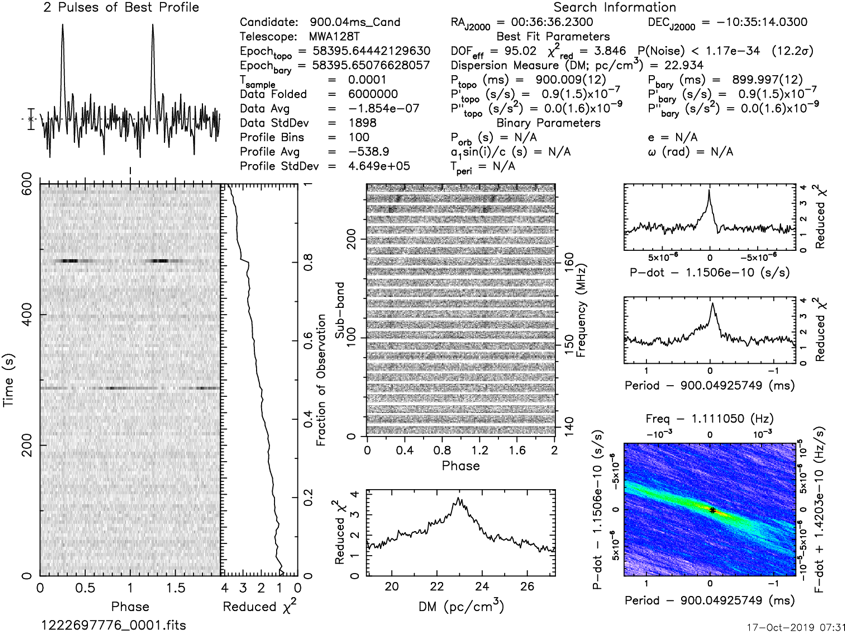}}
 \hfill	
  \subfloat[]{\includegraphics[width=0.5\textwidth]
{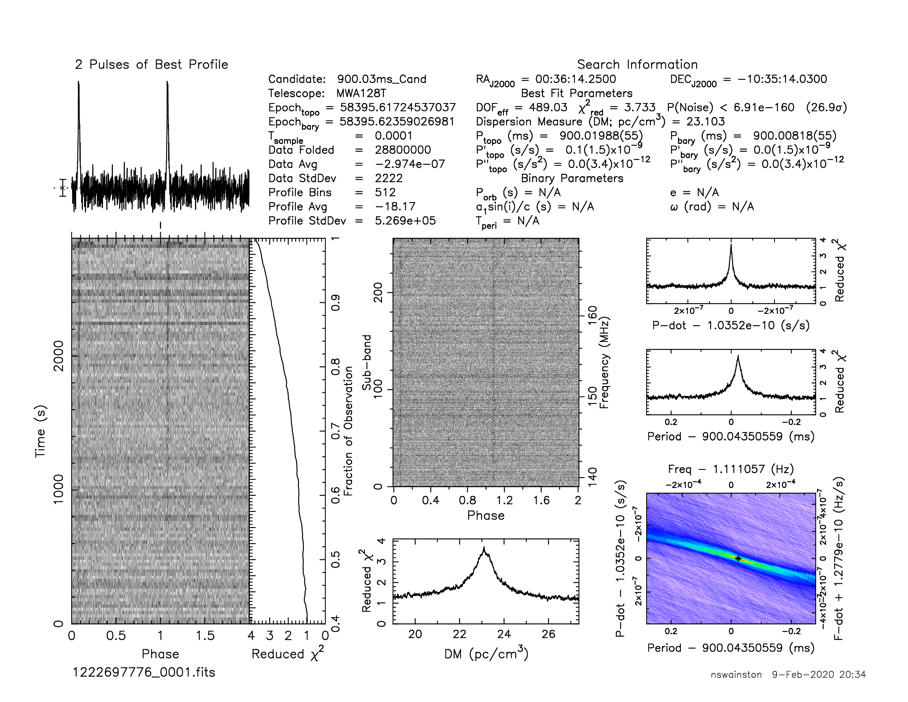}}
 \newline
   \subfloat[]{\includegraphics[width=0.5\textwidth]{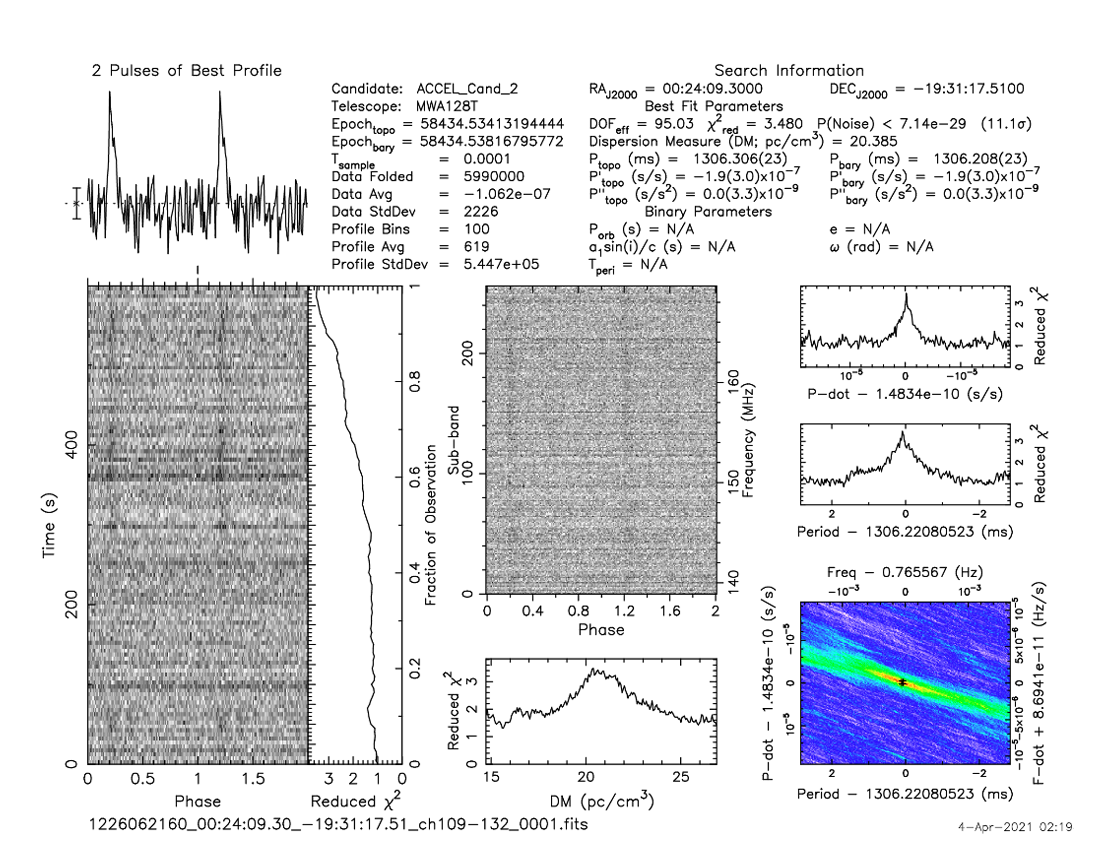}}
 \hfill 	
  \subfloat[]{\includegraphics[width=0.5\textwidth]{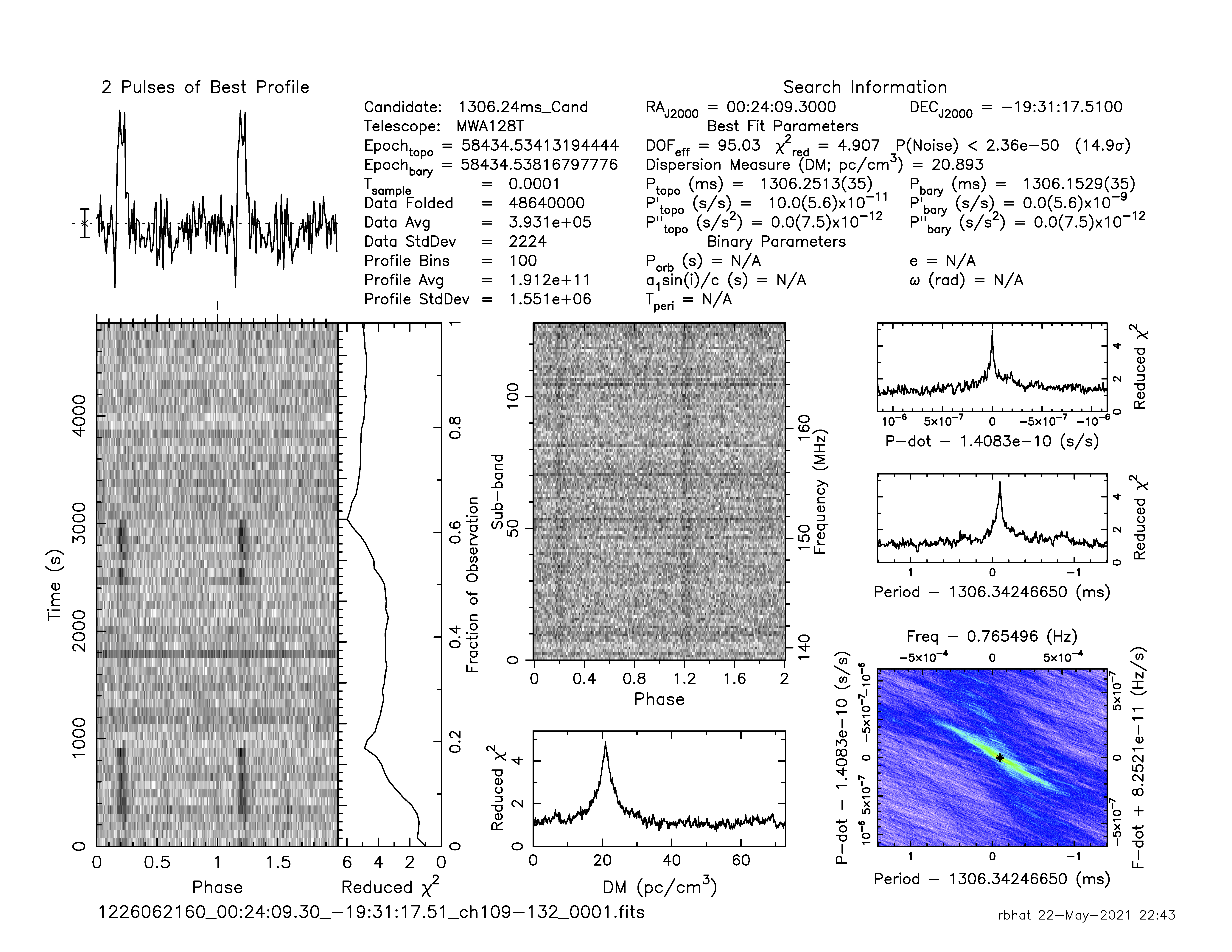}}
 \hfill 	
\caption{Examples of standard PRESTO diagnostic plots of original periodic pulsar candidate detections (left panels), and improved detection plots from follow-up processing for confirmation (right panels). Upper panels are the first pulsar discovered from the SMART, PSR~\psrone, and the lower panels are the second pulsar, PSR~\psrtwo. 
Initial detections are from 10-min observing durations (first-pass processing), while the confirmation ones are from longer durations of the same initial detection observations. 
}
\label{fig:prestoplots}
\end{figure*}

Folding of the candidates is  performed using the {\tt prepfold} tool, which creates the associated candidate files and standard diagnostic plots such as those shown in Fig.~\ref{fig:prestoplots}. Since our pipeline uses the LOTAAS classfier, the folding analysis is carried out using the identical parameter setup as in the LOFAR search pipeline; i.e., 100 pulse phase bins, 256 sub-bands, 120 sub-integrations for $P>10$\,ms, whereas 50 pulse phase bins and 40 sub-integrations for $P<10$\,ms. With this, the folded candidate information can be classified and processed using the ML classifier that we have adopted from the LOFAR search. 

\subsubsection{Single pulse search}
\label{sec:singlepulsesearch}

Single pulse searches have proven to be effective for detecting the class of pulsars that emit sporadically (e.g., RRATs, and giant-pulse emitters such as the Crab).  The basic algorithm involves trialling a range of box car widths, $2^n \, t_{samp}$, where $t_{samp}$ is the sampling time resolution (100\,\us{} for SMART) and $n=0,1,2,\dots, N$, where $N$ corresponds to the maximum width searched \citep[e.g.,][]{cordes-mclaughlin2003}, and detecting `events' that are above a set threshold. It is not computationally demanding, and 
is routinely performed in most pulsar searching. The pipeline has been tested using a SMART observation containing the Crab pulsar, and has also yielded a blind detection of a LOFAR-detected RRAT~J0301+20 \citep{michilli2018}. Integrating this into the processing chain is part of our second-pass search strategy. 

\subsection{Pulsar detection pipeline}
\label{sec:pulsarpipeline}

\subsubsection{ML classification of candidates}
\label{sec:ML_cull}

For each VCS survey pointing ($\sim$610\,\sqdeg,  which is tessellated to $\sim$6000-8000 beams), the processing typically results in $\sim$135,000 candidates.
Scaling for a significantly larger sensitivity (3$\times$) and a larger number of DM trials ($\sim 4\times$) anticipated in full-scale deep searches in the second pass, we may expect over 50 million candidates.
Even for the first pass, as many as 9 million candidates can be expected, extrapolating the rate of candidates requiring scrutiny from the current pipeline.
Indeed, visual inspection of that many candidates is unrealistic, thus necessitating the use of ML classifiers. 

As an initial strategy, we have adopted the ML software that was developed for LOTAAS.
The algorithm used is described in \citet{Lyon2016} and \citet{tan2018a}, and is summarised here.
The classifiers use the statistics of the pulse profile (i.e., mean, variance, skewness and kurtosis) and the DM curve (i.e., S/N vs DM; see Fig.~\ref{fig:prestoplots}).
As described in \citet{tan2018a}, this basic approach is expanded by also calculating the correlation coefficient between each sub-band of the profile, as well as correlation coefficients between each sub-integration and the profile.
In effect, the classifier uses the statistics of correlation coefficient distributions, in addition to the statistics of the profile and the DM curve, in order to classify the periodicity candidates.
Four standard models are used for the regression: (1) decision tree algorithm, (2) multilayer perceptron, (3) probabilistic Bayes classifier, and (4) linear support vector machine.

Even without being trained on MWA data, the software performs reasonably well, with a recall rate of $\sim$83\% for the worst-performing regression model.
While clearly not optimised for a MWA search, it can still provide a significant cull on the number of candidates that require human scrutiny as long as the number of false negatives is kept below an acceptable threshold.
To minimise the false negative rate, we use the provided ``ensemble'' classifier, which labels candidates as positive if at least three models classify them as positive.
Under this criterion, the number of candidates is cut down from the original $\sim$135,000 per pointing down to $\sim$20,000 that require human scrutiny, i.e., an efficiency of $\sim$85\%.
The false negative rate can be lowered further by allowing candidates classified as pulsars by a smaller number of regression models to be passed, but this comes at the cost of also lowering the efficiency.
For the first-pass processing, we find the current arrangement to be an acceptable compromise, but will be implementing an improved ML classifier for the second pass.

Of the remaining $\sim$20,000 candidates per pointing, only a small fraction are true pulsar detections, with the vast majority of candidates consisting of noise and RFI.
Here, we are extending the definition of RFI to include any artefact from the MWA signal path that may result in spurious detections. Owing to the radio-quietness of the 
observatory site, 
such candidates belong almost exclusively to this category, and almost never arise from external sources. 
The most common RFI candidates are those with periods of either 1 second or with a close harmonic relationship (e.g., $0.5\,$s, $2\,$s), relating to the division of data packets by 1-second boundaries.
Such candidates are sufficiently few (and easily identified) that we do not apply any automatic procedure for removing them from our pool of candidates.

\begin{figure*}
    \centering
    \includegraphics[width=0.48\textwidth]{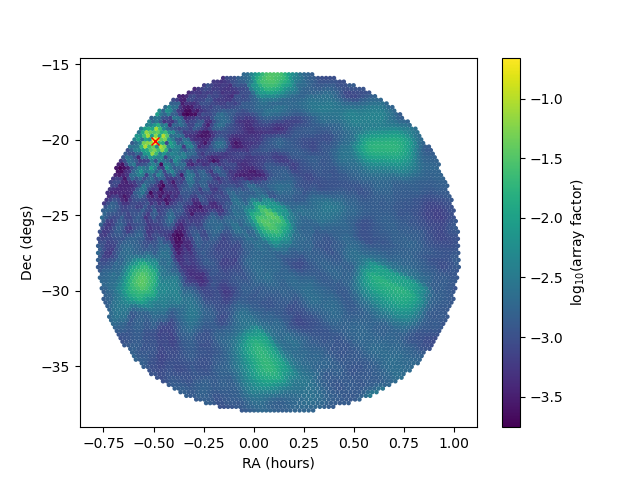}
    \includegraphics[width=0.48\textwidth]{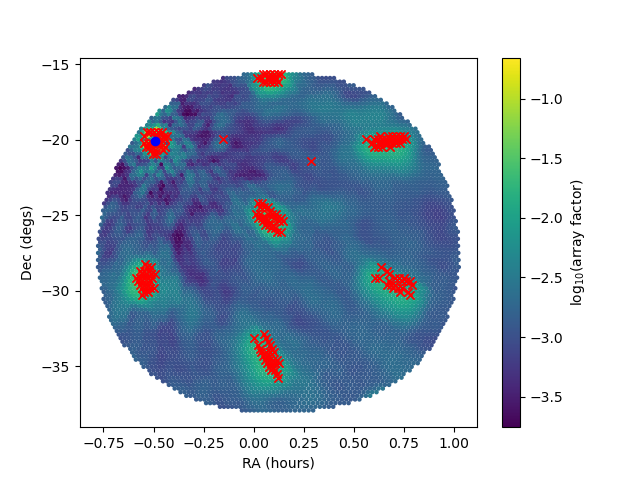}
    \caption{The theoretical array factor (a proxy for sensitivity) of each tied-array beam towards the pulsar B2327$-$20, with the red cross marking the position of the pulsar (left panel) and the beams in which the pulsar
    was detected (right panel). SMART observation 1226062160 was used for the demonstration.}
    \label{fig:array_factor_B2327-20}
\end{figure*}

\subsubsection{Prioritisation and scrutiny of candidates}
\label{sec:prioritisation}

The candidates that survive the initial ML cull are still mostly dominated by noise and RFI detections, with only a small minority being true pulsar detections.
Although all of these candidates are intended ultimately to be visually inspected, we have developed a so-called ``clustering'' algorithm to prioritise which candidates get inspected first, in order to accelerate the detection of sufficiently bright, new pulsars.

The clustering algorithm leverages the fact that the tied-array beam of the MWA's compact configuration is relatively complex, with significant grating lobes located in different parts of the primary beam.
Because the spacing between tied-array pointings is equal to the FWHM of the main lobe of the tied-array beam, any sufficiently bright pulsar will likely be detected in multiple beams.
For instance, Fig. \ref{fig:array_factor_B2327-20} shows a map of multiple detections of PSR~B2327$-$20 superimposed on the theoretical sensitivity of each tied-array beam towards the pulsar, as predicted by the array factor formalism developed for the MWA by \citet{meyers2017}.
Since noise candidates will not be correlated across different beams, prioritising similar candidates that appear in multiple beams dramatically increases the likelihood that candidates representing true astrophysical signals will be inspected first.\footnote{This is counter intuitive to the case of multi-beam surveys with Parkes-like single-dish telescopes, where similar candidates detected in multiple beams across the sky would indicate RFI.}

Candidates are considered similar if
\begin{enumerate}
    \item they appear in at least two adjacent beams,
    \item they have periods within 0.5\% of each other, and
    \item they have DMs within 3\,\dmu{} of each other.
\end{enumerate}

As a demonstration of the usefulness of the clustering algorithm, we show how it would detect PSR~\psrtwo, the second pulsar discovery in the SMART survey \citep{mcsweeney2022}.
In reality, the clustering algorithm was not implemented until \textit{after} PSR~\psrtwo{} was discovered, but it is interesting to note that the first detection (chronologically) of this pulsar was a grating lobe detection (at the time, the candidates were being served up randomly), which motivated the development of the clustering algorithm in the first place.

\begin{figure}
    \centering
    \includegraphics[width=\textwidth]{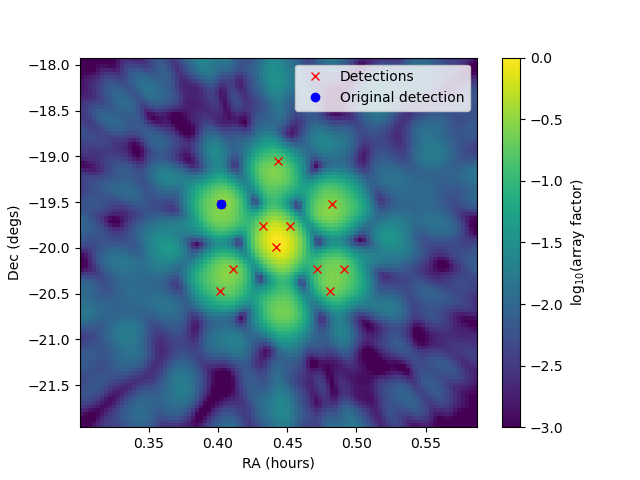}
    \caption{The theoretical array factor (proxy for sensitivity) in the vicinity of PSR~\psrtwo{} for observation 1226062160, assuming a true position (centre of image) derived from GMRT imaging (cf. Paper II for details). Red crosses mark the position of beams in which it was detected, and the blue dot marks the first detection. A single cross may indicate multiple detections with slightly different periods and DMs.}
    \label{fig:array_factor_J0026-1956}
\end{figure}

The final set of detections of PSR~\psrtwo{} is shown in Fig.~\ref{fig:array_factor_J0026-1956}, on a backdrop of the theoretical array factor (a proxy for sensitivity) towards the pulsar assuming that our current best-fit position is correct.
In this case, three of the search beams contained the nominal pulsar position in the main lobe, while several others positioned the pulsar in their respective grating lobes.
All of the displayed detections meet the second and third clustering criteria (similar periods and DMs).
Therefore, any pair of detections in the same or adjacent beams are considered ``clusters'', and if the clustering algorithm was in use when this observation was processed, this pulsar would have been picked up immediately in multiple clusters.

The clustering algorithm offers no advantage for relatively weak pulsars that would be detected only in a single (boresight) tied-array beam.
Therefore, unclustered candidates are not deleted, only deprioritised.

\subsubsection{Human inspection and ranking}
\label{sec:ranking}

Just as the clustering algorithm is a method for prioritising candidates for human inspection, so too is human inspection a method for prioritising candidates for follow-up (see \S\ref{sec:confirmationandfollowup}).
Users are served up candidates one at a time and presented with the candidate's PRESTO diagnostic plots (e.g., Fig.~\ref{fig:prestoplots}).
Each candidate is given an integer rating from 1 to 5, with higher numbers corresponding to a higher confidence that the candidate is a \textit{bona fide} pulsar detection.
Clear pulsar detections are then compared to the ATNF catalogue pulsars to check if it is a known pulsar.
If a detection is unknown, candidates listed in other surveys are then checked using the Pulsar Survey Scraper tool.\footnote{See \url{https://pulsar.cgca-hub.org/}}
If the pulsar is in either the ATNF catalogue or in another survey's candidate list, a note is made against the candidate with the pulsar's name, visible to all other users.

Each candidate can be ranked by multiple users (but users can only rank each candidate once).
A candidate that has been rated by at least four users becomes eligible for follow up, and the list of eligible candidates is ordered by the average rating.

Currently, as the number of users of the system is still relatively small, the rating of candidates is the primary bottleneck in the whole processing chain.
This means that during first-pass processing, interesting candidates have been followed up immediately.
In the future, however, as the number of users performing the task of rating candidates grows, the pool of eligible candidates may grow faster than the rate at which they can be followed up.
However, the above system of candidate prioritisation means that the most interesting candidates are always followed up first.

\subsection{Data management and web-app}
\label{sec:data_management}

The large number of generated candidates, the complex metadata associated with them, and the desire to distribute the tasks of data processing, candidate rating and candidate follow-up, motivated the implementation of a relational database to track the progress of the SMART survey and coordinate processing efforts.
The database, implemented in PostgreSQL, is comprised of a set of tables containing metadata for
\begin{enumerate}
    \item MWA observations (e.g., primary beams, tied-array beams, candidates);
    \item software (e.g., for beamforming, searching, ML classification), including versioning information;
    \item candidate ratings;
    \item pulsars;
    \item users; and
    \item supercomputer facilities.
\end{enumerate}
The users, along with their database access privileges and authentication, are managed by a subset of tables which interface with website front end implemented in Django.
Both the database and the website are hosted by Data Central.\footnote{See \url{https://apps.datacentral.org.au/smart}}

Once an observation has been processed and the candidates have been subjected to the first-pass ML cull (\S\ref{sec:ML_cull}), 
both the metadata of the remaining candidates as well as the candidates themselves (i.e., PRESTO \texttt{.pfd} files and the 
associated diagnostic plots) are uploaded to Data Central.
The uploaded candidates are then available for users to rate via the web interface (\S\ref{sec:ranking}).

As described above, candidates can then be sorted by their average rating, and followed up at will by any authorised user.
Before following up a candidate, the user may ``claim'' it by clicking a button in the candidate list.
This feature is designed to prevent multiple users from following up the same candidate and unnecessarily duplicating effort.
The decentralised design allows members of the SMART collaboration from different research institutions to work through the SMART data set without the need for someone to oversee and coordinate the different groups' activities.


\section{Confirmation and initial follow-up of candidates } \label{sec:confirmationandfollowup}

Confirmation and follow-up of promising pulsar candidates typically relies on multiple re-observations, often requiring a significant amount of telescope time. Fortunately, the SMART survey's unique design, where VCS data are retained (unlike pre-processed beamformed data), offers flexible reprocessing options, allowing us to accelerate important confirmation and follow-up procedures. Furthermore, a substantial amount of archival VCS data (from past projects) are available for a large part of the MWA sky, which can also be suitably exploited for further detection and improved localisation. These features make the SMART survey distinct from other pulsar surveys.

In the following sections we outline the main strategies that are adopted for confirmation and initial follow-up, including: 
reprocessing of the original observation for improved detection; 
performing a dense grid for improved sky localisation; 
and polarimetry via reprocessing the survey observation for full Stokes information and rotation measure (RM) determination. 
Further detailed follow-ups including the use of archival data for timing analysis and imaging for improved localisation are discussed in the companion paper (Paper II). 

\subsection{Improved detection}
\label{sec:improved-detection}

For our ongoing shallow survey, processing the full 80-minute observation itself readily provides an avenue for confirmation. If the source is genuine and a steady periodic emitter, this should result in a three-fold improvement in S/N. The improvement will be reduced if it is an intermittent source; e.g.,  a pulsar with large nulling fraction. Both these possibilities are exemplified in Fig.~\ref{fig:prestoplots}, which shows the original discovery plots along with the improved detections for PSRs~\psrone{} and \psrtwo{}. The full 80-minute observations (42\,TB) containing the original detection can be  processed and searched over a restricted range in $P$ and ${\rm DM}$ using the PRESTO {\tt prepfold} routine. The observations  were also processed using the {\tt pdmp} routine within PSRCHIVE pulsar data processing suite\footnote{See \url{https://sourceforge.net/projects/psrchive/}} \citep{psrchive,vanstraten2012}, to provide a cross-check and a more accurate DM. This is equivalent to undertaking 
a longer observation for confirmation. For many of our candidates, this readily 
provides effective ways of confirming or rejecting a candidate, and eliminates the 
need for securing additional telescope time that most other surveys typically require. \\

While the long dwell time of 4800\,s should in principle result in an increased sensitivity to sporadic or intermittent pulsars, our current first-pass processing does not necessarily benefit from this. 
Given this, the discovery of PSR~\psrtwo{} in the first 10 minutes of observations, a pulsar with long-duration nulls and a nulling fraction of $\sim$77\%, was remarkably fortuitous 
(see Fig.~\ref{fig:prestoplots}).
Details of the discovery, including an analysis of sub-pulse drifting, are reported in \citet{mcsweeney2022}. 
As mentioned therein, this pulsar turned out to have already been reported as a candidate in the GBNCC survey but was blindly (and independently) discovered in 
the SMART survey data. 

\begin{figure*}[t]
\begin{center}
\includegraphics[width=\linewidth]{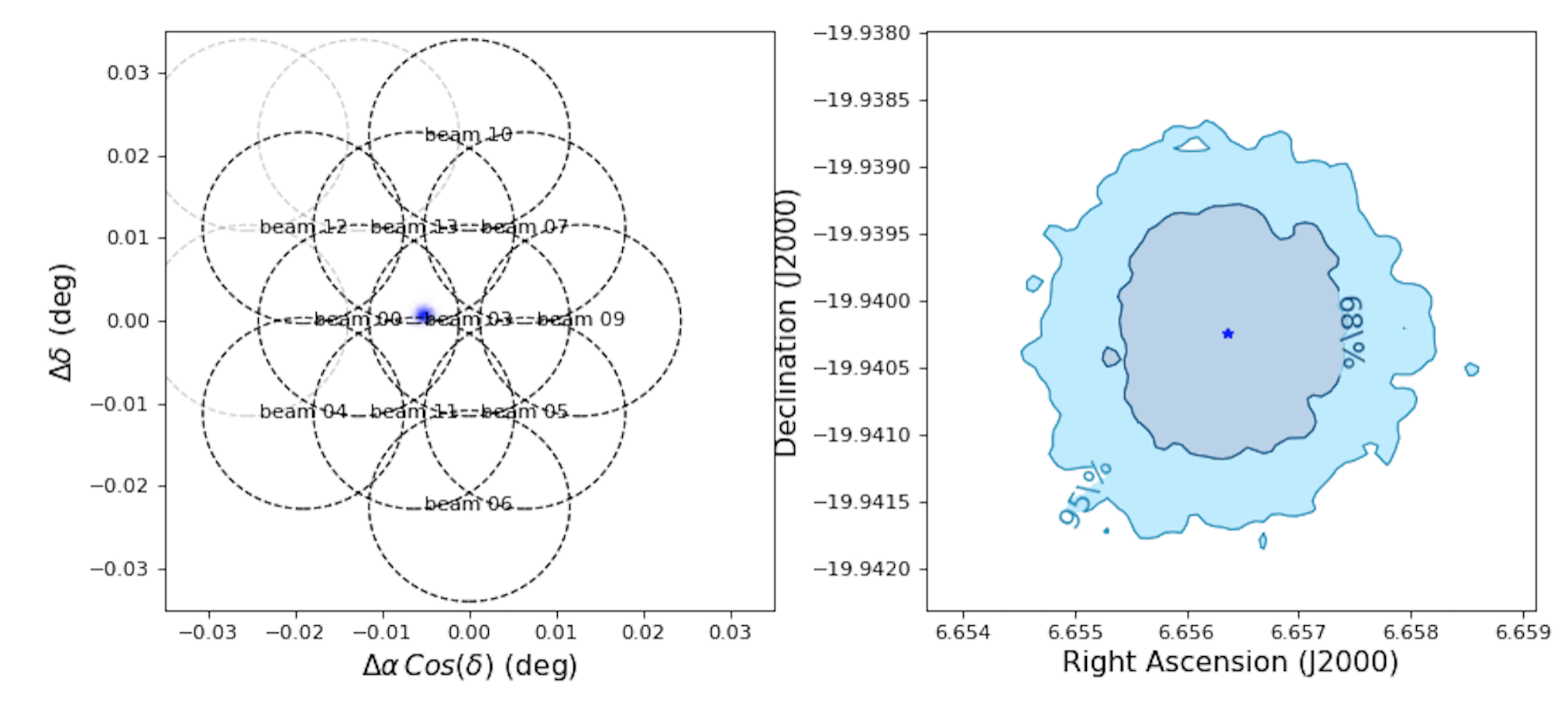}
\caption{MWA localisation of PSR~\psrtwo{} by performing a dense grid around the initial pulsar position from the discovery observation. The source position $\rm (RA, Dec)=(00^h26^m37.5^s, -19\deg 56^\arcmin 24.9^\arcsec)$ is $\approx 32^\arcsec$ offset from uGMRT-determined position 
(cf. Paper II for further details).
Observations were made using the extended MWA array (Phase II, with $\sim$6\,km maximum baseline). The uncertainties in the MWA position is $\sim 12^\arcsec$ (i.e., about one-tenth of the tied-array beam size, shown as dashed circles on the left panel).
}
\label{fig:psrtwo-mwa-localisation}
\end{center}
\end{figure*}

\subsection{Improved positional determination}
\label{sec:improved-position}

As outlined in \S\ref{sec:beamforming}, the tied-array beam size for SMART is $\sim 23^\arcmin$. Therefore a more accurate position is essential both for improved detection (i.e., re-beamforming on a more exact sky position) and to facilitate effective follow-ups with other (and more sensitive) telescopes, particularly at higher frequencies where the beams are narrower, even with single-dish telescopes such as Parkes. This would typically involve making multiple re-observations to form a
grid around the nominal candidate position. The SMART survey design where the sky is densely sampled (at a rate comparable to, or slightly better than, the Nyquist; Fig.~\ref{fig:skytessellation}), allows this to be achieved via reprocessing of the original survey observation, where a dense grid of pointings encompassing the initial position is used for improved positional determination. An example is shown in Fig.~\ref{fig:psrtwo-mwa-localisation} for the case of PSR~\psrtwo. In general, for an initial detection with a modest significance of $\rm S/N \sim 10$, we may expect a positional accuracy $\sim$1-2$^{\prime}$ through this exercise. In practice, archival VCS  data, if available, can also be suitably exploited to progressively further improve the position. In an ideal scenario, where data recorded from all three different configurations are available, an improvement of the order of nearly two orders of magnitude can be achieved through this procedure, as demonstrated in \citet{psrone}.

\subsection{Polarimetry}
\label{sec:polarimetry}

The VCS recording allows the reprocessing of discovery observations to generate full polarimetric beamformed time series, which can be analysed using standard pulsar packages such as DSPSR\footnote{See \url{https://sourceforge.net/projects/dspsr/}} \citep{dspsr} and PSRCHIVE, 
for full Stokes profiles. These beamformed MWA data were obtained using the procedures described in \citet{ord2019} and \citet{xue2019}. The Faraday rotation measure synthesis technique \citep{Brentjens2005} can then be applied to estimate the rotation measure (RM). 

\begin{figure}[t]
\centering 
\includegraphics[width=\textwidth]{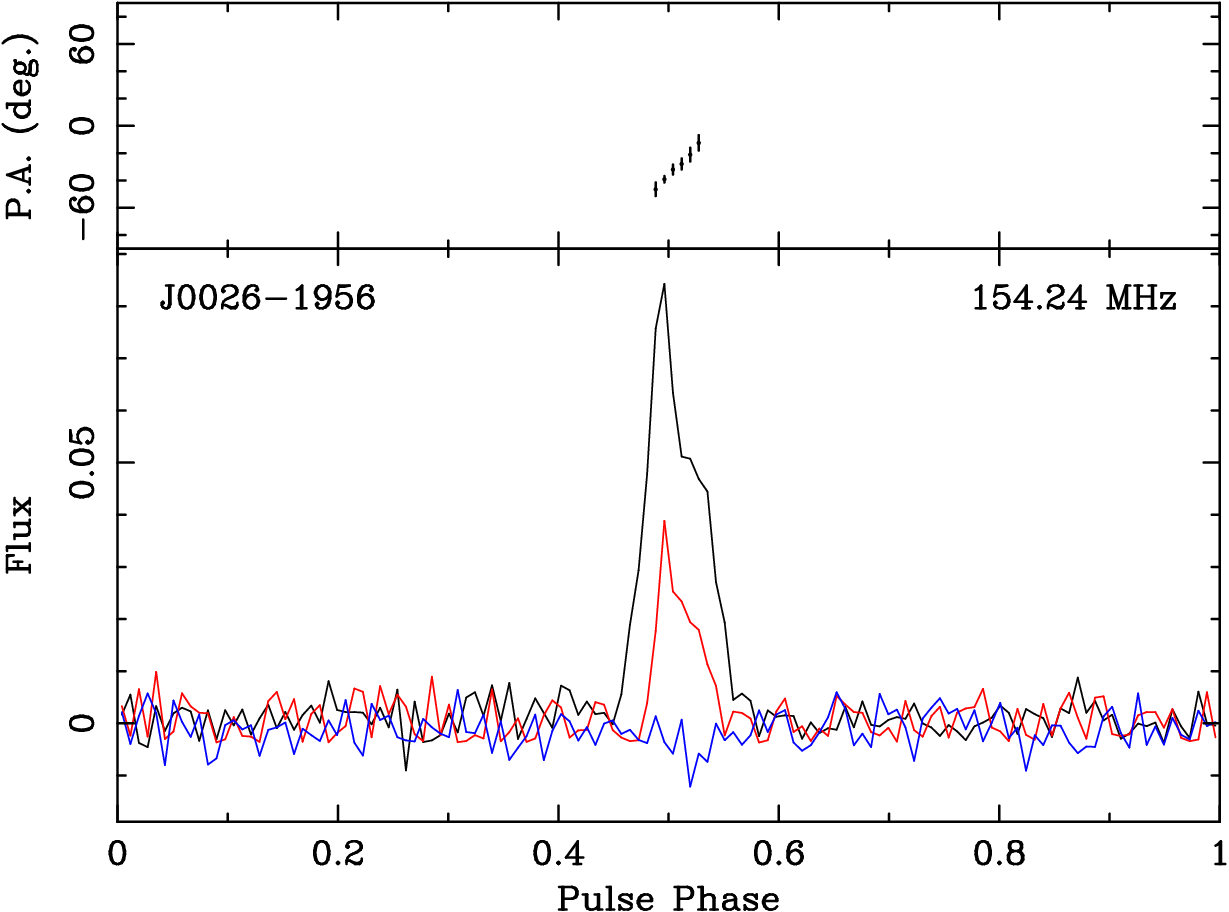}
\caption{Polarimetric profiles of PSR~\psrtwo{} obtained by reprocessing the discovery observation at 155\,MHz. The black, red, and blue curves in the lower panels show the total intensity, linear, and circular polarisation, respectively.
An RM estimate of $ 3.65 \pm 0.09\,\rmu$ was obtained, and the data were corrected for Faraday rotation. 
}
\label{pol}
\end{figure}

As an example, Fig.~\ref{pol} shows polarisation data for pulsar \psrtwo, obtained by reprocessing the original discovery observation. This yielded an RM estimate
of $ 3.65 \pm 0.09 $\,\rmu. After correcting for Faraday rotation, linear and circular polarisation was detected. The pulsar exhibits significant amount of
linear polarisation but only a small amount of circular polarisation. 
We attempted to fit the rotating vector model \citep{Radhakrishnan1969} to the position angle (PA) of the linear polarisation across the on-pulse window, in order to constrain the viewing geometry, $(\alpha, \beta)$, where $\alpha$ is the angle between the magnetic and rotation axes, and $\beta$ is the impact angle of the magnetic axis on the line of sight. In the absence of relativistic effects, the PA curve is expected to be steepest in the center of the pulse profile, with slope $d\psi/d\phi = \sin\alpha / \sin\beta \approx 2.4$, 
where $\psi$ is the PA at phase $\phi$. 

\section{Survey simulations and forecast}
\label{sec:simulations}

The ongoing first-pass processing (i.e., essentially a shallow survey for long-period pulsars) is limited to processing only a fraction (1/8th) of our observation time over coarser (sub-optimal) trial DM values, out to a maximum DM of 250\,\dmu, and to basic periodicity search. In the second pass we will extend this to full 80-min observations and employ more optimal DM steps. Besides a three-fold increase in sensitivity expected for long-period pulsars (by virtue of longer integration times), substantial improvements in sensitivity is also expected to millisecond pulsars via finer DM steps and optimal dedispersion plans to match our 100-$\mu$s/10-kHz resolutions.  
These considerations motivated our simulation analysis to make some meaningful forecast of the expected survey yield, both for long-period pulsars and MSPs, as summarised below. They provide further justification to undertake a full-scale search processing, planned as part of second-pass processing. 

\begin{figure*}[t]
\begin{center}
\vspace{-0.5cm}
\includegraphics[width=\linewidth]{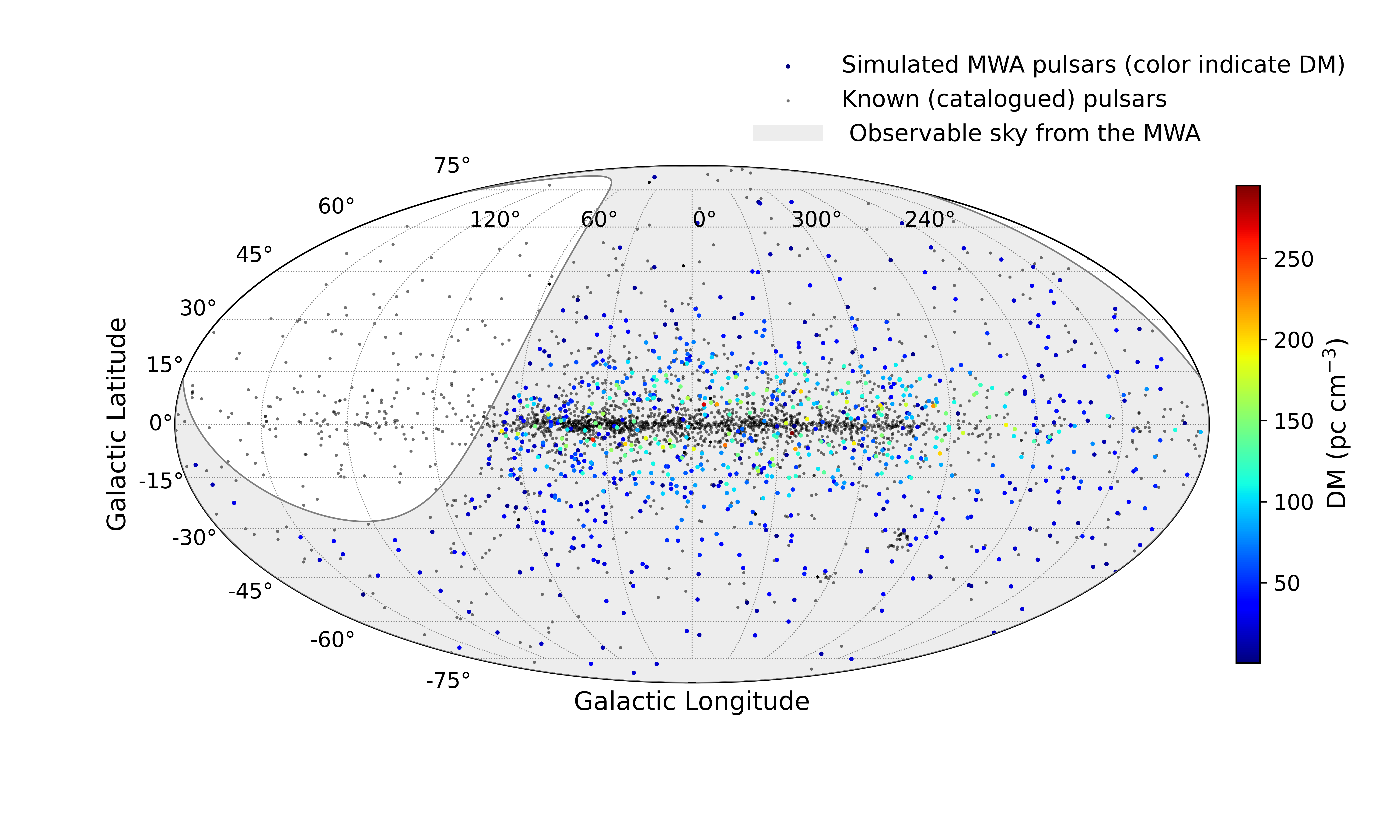}
\vspace{-2.0cm}
\caption{
Simulated pulsars detectable (colour filled circles) in an an all-sky high-time-resolution pulsar search with the MWA in the 140-170 MHz band. 
The shaded region represents the MWA’s visible sky, i.e., the sky south of $+30\deg$ in declination. The black filled circles 
represent known pulsars in the ATNF pulsar catalogue (version 1.67).
The colour scale indicates the DM in units of \dmu. 
}
\label{fig:psrpoppy}
\end{center}
\end{figure*}

\begin{figure*}[t]
\begin{center}
\vspace{-0.5cm}
\includegraphics[width=\linewidth]{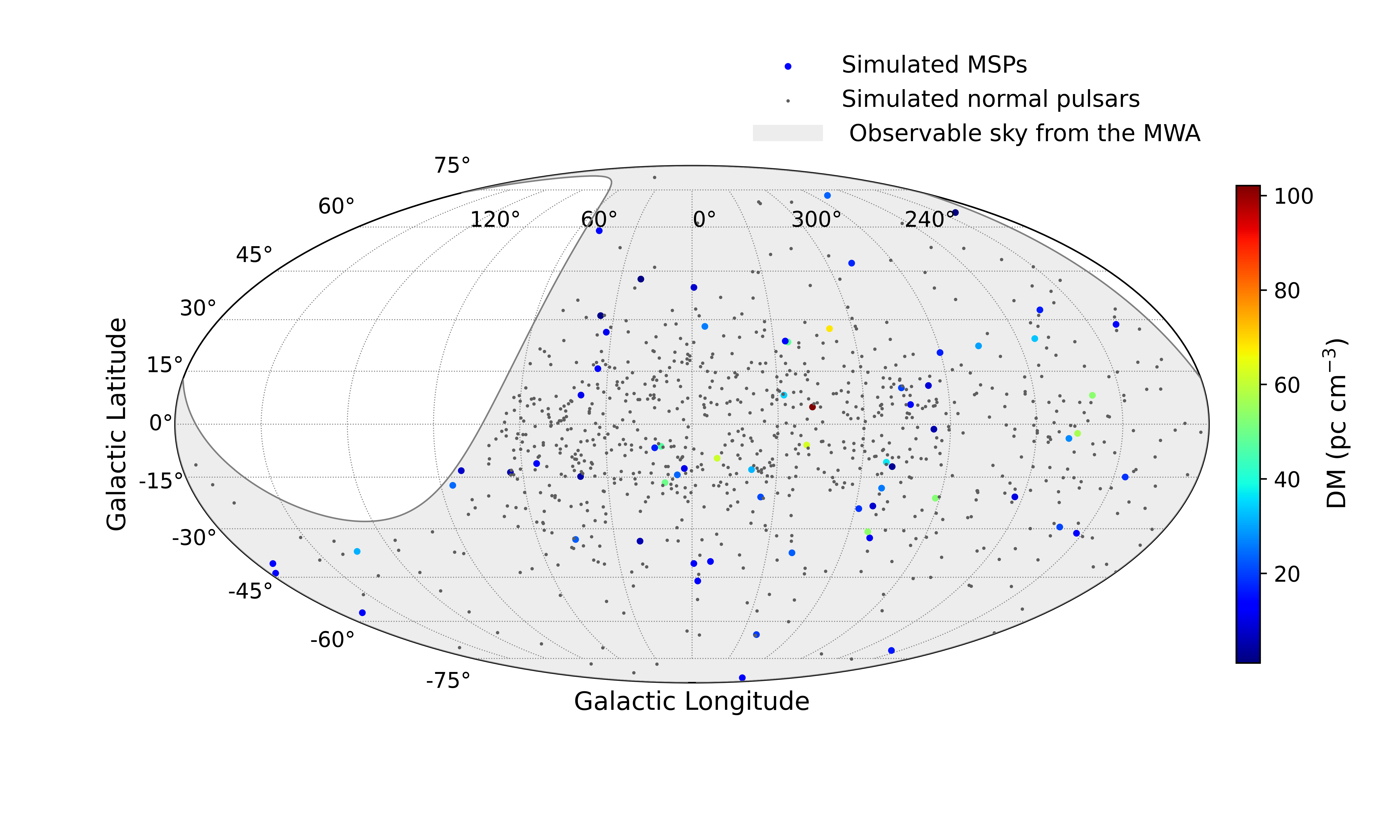}
\vspace{-2.0cm}
\caption{
Simulated pulsars detectable in an all-sky pulsar search with the MWA's 140-170\,MHz band with a dwell time of 4800 seconds. The shaded region represents the MWA’s visible sky, i.e. the sky south of $+30\deg$ in declination. The black filled circles denote the long-period pulsars, whereas millisecond pulsars detectable in high-sensitivity searches (e.g., using the CDMT) are shown as colour filled circles. The colour scale indicates DM in units of \dmu. 
}
\label{fig:psrevolve}
\end{center}
\end{figure*}

\subsection{Long-period pulsars}
\label{sec:longperiodpulsars}

The discovery of two new pulsars from the processing of a small fraction of survey data hints at the potential for many new pulsar
discoveries from a deeper survey that will take advantage of the full 80-min observation. To estimate the survey yield, we have performed survey simulations, using the formalism outlined in \citet{xue2017}. The analysis made use of the popular
simulation package PsrPopPy \citep{Bates2014} that was 
developed from the original pulsar simulation software PSRPOP by \citet{lorimer2006}. The simulations take into account the sky dependence of the system temperatures at  low frequencies  ($T_{sky} \propto \nu^{-2.55}$), as well as the loss 
in the array gain ($G$) expected at large zenith angles, modelled as 
$G(\theta _z) = G_{\rm max}  {\rm cos}(\theta _z)$, where $\theta _z$ is the zenith angle and $G_{\rm max}$ is the gain at $\theta_z=0$. 
We simulated a population of $1.6\times10^{5}$ Galactic canonical pulsars, extrapolated from Parkes Multi-beam Pulsar Survey \citep{manchester2001} detections. The luminosity distribution of the canonical pulsar population follows a log-normal distribution $\langle{\rm log}_{10}L\rangle=-1.1$, $\sigma[{\rm log}_{10}L]=0.9$ , where $L$ is the radio psuedo-luminosity in units of \lmu{} \citep{fauchergiguere2006}. The Galactic radial density distribution follows the \cite{yusifov2004} model.
With the caveat that our understanding of the pulsar luminosity function and beaming fraction is limited, we project the deep survey to reach a limiting sensitivity of $\sim$2-3\,\mJy, with a potential net yield of $310 \pm 100$ new pulsar discoveries (see Fig.~\ref{fig:psrpoppy}). 
This projection mainly applies to the population of long-period pulsars and does not account for other classes of pulsars such as sporadic emitters (e.g., RRATs), or millisecond and binary pulsars, whose populations are hard to model or simulate. 

Assuming an isotropic distribution of our simulated local pulsar population ($\rm DM \lesssim 250$\,\dmu), and scaling for the current (first-pass) search sensitivity (i.e. one-third of the deep pass sensitivity), and the fraction of data for which the candidate scrutiny has been completed ($\sim$5\%), we may expect $\sim$3-5 pulsars. The detection rate at this early stage of 
SMART thus appears to be in line with this general expectation. 
While this may seem fortuitous, the unique advantages of the SMART pulsar survey, especially the accessibility to the southern hemisphere, the radio-quiet environment, and the survey parameters (e.g., long dwell times and high time/frequency resolutions), offer excellent prospects for new pulsar discoveries, provided the substantial processing challenges can be addressed. 

\subsection{Millisecond pulsars}
\label{sec:millisecondpulsars}

Even though the detection sensitivity to MSPs is significantly reduced in our current shallow pass of the survey (owing to the use of coarse or sub-optimal DM step sizes; see Fig. 2), the second-pass processing, where we plan to employ more optimal  DM searches with a finer step-size in DMs, is expected to yield a substantial improvement in sensitivity, particularly at low to moderate DMs, out to $\lesssim$50\,\dmu. At DMs $\gtrsim$70\,\dmu, and especially in regions near the Galactic plane and toward the centre, scatter broadening is expected to result in sensitivity degradation, given the strong frequency dependence (pulse broadening time, $\tau _d \propto \nu^{-3.9}$; cf. \citealp{bhat2004}), due to which $\tau _d \gtrsim $10\,ms, which, for millisecond pulsars, can be a substantial fraction of the rotation period. Using PsrPoPy, we simulated a population of $3\times10^{4}$ MSPs with $P$ and ${\rm DM}$ distributions essentially derived from the HTRU intermediate latitude pulsar survey \citep{levin2013}, and with a luminosity limit of $L_{1400} \sim 0.2\,\lmu$. This corresponds to a limiting flux density $\sim$10\,\mJy{} at 150\,MHz, assuming a spectral index of $\alpha = -1.8$ (and a distance of $\sim$1\,kpc), and thus in principle detectable provided there is no significant degradation from dispersive smearing or temporal broadening from scattering.

As with the population of long-period pulsars, this analysis accounted for the sky dependence of \Tsky, non-uniformity in the array gain, and strong frequency scaling of scattering, 
which is especially important for MSPs. 
For example, using some preliminary dedispersion plan estimates for the second round of processing (i.e., the deep search), where we assume a typical plan would involve DM steps of 0.01\,\dmu{} up to 54\,\dmu{} and 0.02\,\dmu{} out to 107\,\dmu, our simulations predict 55 detectable MSPs above our detection threshold, and hence $\sim$15 new MSP discoveries.
However, a substantial increase is forecast in simulations that closely emulate the higher sensitivity attainable through more optimal searches that make use of \emph{coherent dispersion measure trials} (CDMT), which is equivalent to the use of finer DM steps of 0.002\,\dmu, and will limit residual DM smearing to $\sim 150\,\us$ (comparable to $\sim100\,\us$ native resolution of the VCS). In essence, this means that full-scale, high-sensitivity searches employing the implementation of CDMT, if feasible for SMART, can potentially lead to the discovery of as many as $\sim$30 MSPs.

The simulated population of $\sim$70 MSPs, along with the simulated population of long-period pulsars (see \S\ref{sec:longperiodpulsars}), is shown in Fig.~\ref{fig:psrevolve}. Our simulation analysis did {\it not} consider a large population of MSPs discovered in recent (and highly successful) Fermi-directed targeted searches \citep[and references therein]{deneva2021}. Even so, the detectable population of MSPs is almost twice the currently known population within $\rm DM \lesssim 100$\,\dmu, 
which means a net MSP yield that is competitive to that from the highly successful Parkes HTRU survey. Indeed, as evident from Fig.~\ref{fig:psrevolve}, the detectable population of MSPs is limited to $\rm DM \lesssim 70$\,\dmu, which is reconcilable given the expected pulse broadening times of $\tau _d \gtrsim $10\,ms toward such moderate-DM pulsars at the low frequencies of the MWA \citep[e.g.,][]{kirsten2019}. 
Consequently, the vast majority of MSPs discovered will likely be suitable for high-precision timing applications such as pulsar timing arrays.

\section{Future processing plans } \label{sec:discussionandfutureplans}

The planned second-pass survey  will extend the processing to the full 80-min observations and carry out more optimal searches in the DM parameter space, while incorporating  searches for both long-period pulsars and millisecond pulsars. As such, the long dwell times of SMART (4800\,s) can be exploited to search for pulsars with very long periods, like those discovered by LOFAR and MeerKAT \citep{tan2018b,caleb2022}, and provide increased sensitivity to objects that emit intermittently, e.g. pulsars with long null duraions such as PSR~\psrtwo{} \citep{mcsweeney2022}. In addition, the adopted strategy to archive recorded voltages offers additional avenues for future processing; e.g. searches for millisecond pulsars through the application of novel hybrid dedispersion approaches that involve the use of coherent dispersion measure trials (CDMT), which was demonstrated by the LOFAR through the discovery of PSR~\cdmtpsr{} \citep{bassa2017}. 
Below we outline our processing plans and strategies in the near-term and highlight some of the computational challenges and other considerations in planning this second-pass processing. 

\subsection{Beamforming and sensitivity optimisation}\label{sec:bfsensplans}
As discussed earlier in \S\ref{sec:effective_dwell_sens}, the tied-array beamforming strategy warrants some more careful thought in order to maximise sensitivity while also reducing needless processing. Inevitably, this produces an uneven sensitivity threshold across the sky due to both primary beam pointing effects and effective dwell time. These considerations are also important when estimating survey-wide statistics. We are formulating a more efficient beamforming scheme that takes into account these technical details, which will be presented in a subsequent paper detailing the second-pass survey processing.

\subsection{Dedispersion planning and RFI mitigation strategies}\label{sec:dedisprfiplans}
For the first-pass survey processing described in this paper, the dedispersion plan outlined in Table~\ref{tab:ddplanfirst} is adequate for all observations.
In contrast, a slightly more sophisticated plan may be required for the second-pass processing to accommodate the eight-fold increase in observation length and to provide increased sensitivity to shorter-period pulsars.
We are actively developing a sensible strategy that balances our sensitivity goals and the relatively large computational costs associated 
with dedispersing MWA VCS data, especially since we would essentially be producing $\sim$10$\times$ as many DM trials.

In addition to revisiting the dedispersion plan, we will also incorporate a more careful approach to excising or mitigating RFI (both periodic and impulsive). 
The observatory site
is exceptionally RFI-quiet (owing to the geographical location and radio-quiet zone status), hence the first-pass processing did not include any active RFI mitigation other than what is naturally gained by forming TABs (where off-axis RFI is ``phased out''). 
We are currently examining the periodic RFI environment by processing observations taken throughout the SMART observing semesters and using a standard PRESTO-based approach to find bright, common terrestrial signals by searching for periodic ``candidates'' in the zero-DM topocentric time series data.
Once we collect this information, we will apply the masks (after appropriate barycentric corrections are made) during the periodicity search pipeline. 
Additionally, there can occasionally be bursts of narrow-band interference (e.g., air-craft and satellites in TAB grating lobes) that could severely affect our data quality for short periods of time.
There are several software pre-processing solutions to this kind of RFI \citep[e.g.,][]{zdm,zdmf,iqrm}, which we will explore in parallel to the periodic RFI mitigation strategies.
Empirically, VCS data are remarkably clear of impulsive/narrowband RFI 
in the SMART observing band,
and data excision is $\ll$10\% for a typical observation.

\subsection{Searches for long-period pulsars and sporadic emitters}
\label{sec:deepersearches}

The long dwell times of SMART make it particularly amenable to the application of fast-folding algorithms that offer significantly higher sensitivity to pulsars with rotation periods $\gtrsim$10\,s \citep[e.g.,][]{morello2020}. Such slow-spinning pulsars are likely to be near the 
radio emission
`death lines' so 
can be invaluable in gaining useful insights into the intricacies of the pulsar radio
emission process. Recent applications of this algorithm in Parkes and Arecibo searches have led to the discoveries of pulsars with $P >$10\,s and \citep{morello2020}, or very weak pulsars ($S_{1400}\sim10$\,$\mu$Jy) with a $\sim$2\% duty cycle \citep{parent2018}. These, and other recent discoveries such as 
a 76-s pulsar with MeerKAT \citep{caleb2022}, provide a strong motivation for undertaking fast-folding searches. The low levels of RFI at the 
observatory
site are particularly advantageous for this. 

The SMART survey dwell time is substantially longer than those of previous-generation southern-sky surveys, particularly at high-$|b|$ parts of the sky, where it is 20 times longer than the HTRU survey \citep{keith2010} and 40 times longer than the southern pulsar survey \citep{70cm}.
It is also 40 times longer than that of the ongoing GBNCC survey \citep{stovall2014} that covers the sky north of $-55\deg$ in declination (Table~\ref{tab:surveys}). 
Considering this, detection prospects are promising, especially given the $\sim$2-3\,\mJy{} limiting sensitivity that the SMART can attain for long-period pulsars (\S\ref{sec:surveysensitivity}) and negligible degradation in signal strength due to dispersion and pulse broadening effects.

As described earlier,
the long dwell times also increase the search sensitivity to objects that emit sporadically, such as RRATs and giant-pulse emitters (e.g., the Crab pulsar), which can be more effectively detected by searching for individual dispersed pulses, and will be part of the second-pass processing. 

\subsection{Searches for binary and millisecond pulsars} \label{sec:acceleration}

The long dwell times, and high time and frequency resolutions, of the SMART can also be exploited, in principle, to search for binary and millisecond pulsars. However, a full-scale acceleration search can be prohibitively expensive at the low frequencies, given the very large number of DM and acceleration trials that are required (e.g., typically $\sim 10^4$ up to 250\,\dmu, and $\sim 2400$ across $\pm 100\,\accu$). Compared to the HTRU-south low-latitude survey, which has been successful in finding such systems \citep[e.g.,][]{cameron2020}, the cost of searching SMART data can be more than an order of magnitude greater. 
The successful detection of several MSPs and the double pulsar in our initial census (cf. Paper II) makes such searches worthwhile. 

An inherent limitation in the searches for such short-period pulsars is the significant degradation in sensitivity due to substantial dispersion smearing (relative to rotation periods) despite our 10-kHz channels. Fortunately, this can be alleviated by using CDMT-based searches \citep{bassa2017}. Recording in 24$\times$1.28-MHz channels makes the SMART data highly amenable to the application of CDMT searches, and can result in a substantial increase in detection sensitivity to short-period millisecond pulsars.
Integration of this novel method, and benchmarking on prospective HPC clusters with significant computational resources (e.g., Pawsey's emerging Setonix cluster) is also part of our future processing plans, although a full-scale processing may have to await access to sufficient computational resources.
We are also exploring publicly available, GPU-enabled Fourier domain acceleration search software \citep[e.g., AstroAccelerate;][]{astroaccelerate} as a drop-in replacement for PRESTO's CPU-based {\tt accelsearch}.

Regardless, the high cost of such computationally-intensive searches will likely necessitate a multi-pass processing strategy; for instance, an initial pass involving acceleration searches, but limited to a modest number of acceleration trials (e.g., $\sim$150 to cover $\pm 6\,\accu$), thereby retaining sensitivity to short-period objects ($P \lesssim 10$\,ms) but with the binary orbital period, $P_b \gtrsim 5$\,days (i.e., with low-mass white-dwarf type companions). Full-scale acceleration searches that target binary systems such as PSR~J0737$-$3039 or PSR~J1757$-$1854 with $P_b \lesssim 5 $\,hr (i.e., requiring $\sim$2400 trials spanning across $\pm 100\,\accu$) are hence deferred to the longer-term future. Such searches will be primarily limited to the regions around the Galactic plane, at least initially, thus processing only a fraction of the SMART data (e.g., sky within $|b|\lesssim5\deg$). Such a multi-pass strategy is also motivated by the demonstrated success of HTRU-south, which has led to the discovery of exotic systems such as PSR~\hitrun{} \citep{cameron2018} and wide-orbit double neutron-star system \citep{sengar2022}. In any case, notwithstanding the high computational cost, the high-profile scientific applications of such rare systems make similar full-scale acceleration searches scientifically compelling for the SMART data. The long-term scientific dividends of such systems are vividly demonstrated by \citet{kramer2021} through the 16-year timing analysis of the double pulsar, enabling the most stringent tests of general relativity and alternative theories of gravity.



\section{Summary and conclusions }
\label{sec:summary} 

With its novel features such as voltage recording and long dwell times, and access to the pristine 
radio-quiet environment in the southern hemisphere, the SMART survey is well positioned to play an impactful role in the exploration of the southern, low-frequency sky for pulsar surveys and science. Since the MWA is a precursor for SKA-Low, the SMART survey will also serve as an important preparatory step for pulsar surveys planned with SKA-Low. Additionally, it will map out the southern sky for low-frequency detections of many pulsars that were originally discovered at frequencies $\gtrsim$400 MHz. 

The survey is enabled by the advent of the Phase II upgrade of the MWA, the compact configuration of which offers an enormous gain in the beamforming and processing cost, thereby making large all-sky pulsar surveys tractable with large-FoV interferomtric arrays such as the MWA. The combination of voltage recording and the FoV brings a survey efficiency of $\sim$450\,\sqdegphr, but at the expense of large data rates of 28\,\TBhr. Consequently, $\sim$3\,PB of (VCS mode) data for the full survey and significant processing costs.

Due to the substantial computational cost involved in searching at low frequencies, the processing is undertaken in multiple passes. 
In the ongoing first-pass processing, 10\,min of data from each observation are processed in 2358 trial DMs, out to a maximum DM of 250\,\dmu, thereby reaching about one-third of the sensitivity that will eventually be attainable in full observation processing.

The voltage recording strategy adopted for the SMART survey enables a multitude of avenues for follow-ups and confirmations, including improved detection, initial polarimetry and arcminute-level positional determination -- all by reprocessing the original observation and, where possible, also archival VCS data.
This also facilitates timely follow-up studies using more sensitive telescopes such as Parkes and uGMRT that operate at frequencies $\gtrsim$300 MHz.

With the recent development of a {\it web-app} for facilitating efficient scrutiny of candidate analysis,
including classification and ranking for identifying promising ones to follow-up, 
we anticipate the discovery rate to increase in the coming years. As software tools mature and the search pipelines are expanded to include acceleration trials and fast-folding based algorithms, and additional computational resources become available, it will become possible to extend the processing to include searches for binary and millisecond pulsars, and those with very long periods or even sporadic emitters. 
Our simulation analysis forecasts a survey yield of $\sim$300 long-period pulsars and $\sim$30 millisecond pulsars by the completion of full processing. 
The SMART survey data will serve as a complete digital record of the low-frequency southern sky, and an important reference for even more ambitious surveys planned with the SKA-Low.

\begin{acknowledgement}

We thank an anonymous referee for several useful comments that helped to improve the content and presentation of this paper. 
The scientific work made use of 
Inyarrimanha Ilgari Bundara, the CSIRO Murchison Radio-astronomy Observatory. 
We acknowledge the Wajarri Yamaji people as the traditional owners of the Observatory site. This work was supported by resources provided by the Pawsey Supercomputing Centre with funding from the Australian Government and the Government of Western Australia.
This work was also supported by resources awarded under Astronomy Australia Ltd's ASTAC merit allocation scheme on the OzSTAR national facility at the Swinburne University of Technology. The OzSTAR program receives funding in part from the Astronomy National Collaborative Research Infrastructure Strategy (NCRIS) allocation provided by the Australian Government.
The development of SMART webapp was facilitated by the software support scheme of ADACS.
We thank Simon O'Toole for help with the migration of webapp to Data Central.  
The GMRT is run by the National Centre for Radio Astrophysics
of the Tata Institute of Fundamental Research, India. 
The Parkes (\emph{Murriyang}) radio telescope is part of the Australia Telescope National Facility which is funded by the Australian Government for operation as a National Facility managed by CSIRO.
We thank L. Levin for help with the simulation analysis of MSPs. 

\textit{Software}: We acknowledge the use of the following software/packages for this work: CASA \citep{casa}, DSPSR \citep{dspsr_ascl, dspsr}, PRESTO \citep{presto, presto_ascl}, PSRCHIVE \citep{psrchive,psrchive_ascl}, Tempo2 \citep{tempo2_1, tempo2_ascl}, PINT \citep{pint_ascl, pint}, PsrPopPy \citep{Bates2014}, Nextflow \citep{nextflow}.

\end{acknowledgement}


\bibliography{ms}

\end{document}